\documentclass[aps,floats,showpacs,amssymb,tightenlines]{revtex4}

\usepackage{amsmath}
\usepackage{amsfonts}
\usepackage{amscd}
\usepackage{epsfig}
\usepackage{amssymb}
\usepackage{tabularx}

\def\e{\epsilon}

\def\w{\omega}

\def\p{\partial}

\def\I{{\cal I}}

\def\po{\psi_{\w_o}}
\def\pe{\psi_{\w_o}^{(j=1)}{}}

\begin{document}
\title{The initial value problem for linearized gravitational perturbations of the Schwarzschild naked singularity}

\author{Gustavo Dotti and Reinaldo J. Gleiser}
\affiliation {Facultad de Matem\'atica, Astronom\'{\i}a y
F\'{\i}sica (FaMAF), Universidad Nacional de C\'ordoba. Ciudad
Universitaria, (5000) C\'ordoba,\ Argentina}

\begin{abstract}
The coupled equations for the scalar modes of the linearized Einstein equations around Schwarzschild's spacetime were reduced 
by Zerilli to a 1+1 wave equation $\p^2 \Psi_z / \p t^2 + {\cal H} \Psi_z =0$,
where $ {\cal H} = -\p ^2 / \p x^2 + V(x)$ is the Zerilli ``Hamiltonian", 
$x$ the tortoise radial coordinate. 
From its definition, for smooth metric perturbations 
the field $\Psi_z$ is singular at $r_s=-6M/(\ell-1)(\ell+2)$, with $\ell$ the 
mode harmonic number. The equation $\Psi_z$ obeys is also singular, 
since $V$ has a second order pole at $r_s$.
This is irrelevant to the black hole exterior stability problem,
where $r>2M>0$, and $r_s <0$, but it introduces a non trivial problem in the naked singular case where $M<0$, 
then $r_s>0$,
and the singularity appears in the relevant range of $r$ ( $0 < r< \infty$).
We solve this problem by developing a new approach to the evolution of 
the even mode, based on a {\em new gauge invariant function}, $\hat \Psi$, that is a regular
 function of the metric perturbation
{\em for any value of $M$}.
The relation of $\hat \Psi$  to
 $\Psi_z$ is provided by an intertwiner operator. 
The spatial pieces of the $1+1$ wave equations
that $\hat \Psi$  and  $\Psi_z$  obey are related as a supersymmetric pair of quantum hamiltonians
${\cal H}$ and $\hat {\cal H}$. For $M<0$, 
 $\hat {\cal H}$ has a regular potential and a unique self-adjoint extension in 
a domain ${\cal D}$   
defined by a physically motivated boundary condition at $r=0$. This allows 
to address the issue of evolution of gravitational perturbations 
 in this non globally hyperbolic background.
 This formulation is used to complete the proof of the linear instability of
the  Schwarzschild naked singularity, by showing that a previously found unstable mode
belongs to a complete basis of $\hat {\cal H}$ in ${\cal D}$,
and thus is excitable by generic initial data. This is further illustrated by numerically solving the 
linearized equations for suitably chosen initial data. 
 \end{abstract}


\maketitle

\section{Introduction}
The linear stability under gravitational perturbations of the negative mass Schwarzschild
spacetime was first considered in \cite{ghi}, where a proof of stability for the
vector (or odd) modes is given. For the scalar (even) modes, reconsidered in \cite{dg},  the problem is far more subtle,
because the behaviour of the Zerilli potential $V_z$ \cite{zerilli,ki} at $x=0$ (which corresponds to
the $r=0$ Schwarzschild singularity) implies
a one parameter ambiguity \cite{ghi} in boundary conditions at this point (parameterized by $\theta \in S^1$,
 see equation 
(\ref{sae}) below), and
  also because   $V_z$ has 
a second order pole at $r=r_s:=-6M/(\ell-1)(\ell+2)$  which  falls within the domain of interest 
for negative $M$. None of these problems
are present in the positive mass case, for which the relevant range is
$r > 2M$ (mapped over $-\infty < x < \infty$),  and $r_s <0$.

The ambiguity in boundary conditions at $x=0$ was addressed to in \cite{ghi,dg}, where it was shown 
 that   $\Psi_z \sim x^{1/2} $   as $x \to 0^+$ is to be selected in order  that the first
 order corrections to the Riemann tensor algebraic invariants do not diverge faster than
their zeroth order piece as the singularity is approached, 
a  natural requirement if one wants the first order
formalism to provide approximate solution of Einstein's equations that can be consistently interpreted as
 arbitrarily small perturbations of the unperturbed metric. This choice also 
selects perturbations with finite energy, using the energy notion obtained by going to
second order perturbation theory \cite{ghi,dg}.   The singularity of $V_z$ at 
$r=r_s$ is called a ``kinematic"  in \cite{dg}, because it is due 
to the fact that, as 
defined, the Zerilli function $\Psi_z$ has a simple pole
at this point for generic smooth gravitational perturbations 
(see \cite{dg} and  Lemma 1, equation (\ref{crc}) below.) In the Zerilli 
formulation \cite{zerilli, ki}, the initial value problem (IVP) for linearized gravity around the $M<0$ Schwarzschild
spacetime can then be posed as follows: 
given $\Psi_z(t=0,x), \dot \Psi_z(t=0,x)$ defined for $x>0$, 
both satisfying (\ref{crc}) 
and vanishing as $x^{1/2}$ -or faster- for $x \to 0^+$, find the unique $\Psi(t,x)$ obeying the 
singular equation equation (\ref{zerilli})-(\ref{tor}) in the half space $x>0$, and giving this 
data for $t=0$. \\
The purpose of this paper is to solve this rather bizarre IVP.
The exterior black hole
 ($r>2M>0$) Zerilli equation is entirely free of difficulties, 
it is a  $1+1$ wave equation in a {\em complete} Minkowskian space (the horizon 
lying at the tortoise coordinate value $x=-\infty$), with a smooth  potential, and 
initial data can be evolved by ${\cal H}$ mode  expansion. 
The difficulties for the $M<0$ case cannot be overcome 
 by introducing alternative radial variables or integrating factors, which
can be easily seen to merely move the singularity from the coefficients of the differential equation to the measure
that makes its radial piece self adjoint.  
Solving the IVP for $M<0$ allows us to complete the proof in \cite{dg} that
the negative mass Schwarzschild spacetime is unstable under linear gravitational perturbations, as part of a program 
to study  the linear stability 
of the most notable nakedly singular solutions of Einstein's equation \cite{dgp,dgsv},  
in connection to  cosmic censorship. Unstable (exponentially growing in time)  modes were not only found for 
 the negative mass Schwarzschild spacetime \cite{dg}, but also for the $|Q|>M$ Reissner N\"ordstrom 
 and the $|J| >M^2$ Kerr naked singularities \cite{dgp,dgsv}. The instability for the negative mass Schwarzschild -(A)dS 
and the negative mass Reissner-N\"ordstrom spacetimes were proved in \cite{cardo}.
The unstable smooth solutions of the $M<0$ Schwarzschild 
 linearized Einstein equations in \cite{dg}, satisfy the desired
boundary condition at $r=0$, and decay exponentially  for large $r$.
It is  argued  in \cite{dg} 
that they can be excited  by initial data  compactly supported
away from $r=0$, but this can not be proved if we do not know how to evolve initial data.
In this paper we show how the IVP for  even
perturbations on a negative mass Schwarzschild spacetime can be solved by using the technique of intertwining operators
 (see \cite{price} and references therein).
An intertwining operator $\I= \p/\p x + g(x)$ is constructed such that for
 regular metric perturbations $\hat \Psi := \I \Psi_z $
is smooth and belongs to
$L^2((0,\infty),dx)$.
$\hat \Psi$ satisfies a Zerilli like equation $0= [\p^2 /\p t^2 +\hat {\cal H} ] \hat \Psi$, $\hat {\cal H} :=
 - \p^2/\p x^2 + \hat V$,
with a potential $\hat V$ that is  free of  singularities and such that $\hat {\cal H}$ has  {\em a unique}
self adjoint extension in a domain ${\cal D} \subset L^2((0,\infty),dx)$, that corresponds precisely
 to our physically motivated choice of boundary condition at $x=0$.

These two key differences with the standard Zerilli approach allow us to give a
 comprehensive answer to the linear stability problem
of $M<0$ Schwarzschild spacetime, as we can show that physically sensible initial
 data supported away from the singularity generically
excites the unstable modes found in \cite{dg}. As is shown in Section \ref{num}, 
this is {\em not} related to the $x=0$ boundary
conditions; if a perturbation is initially supported away
 from the singularity, {\em the unstable modes are excited before the excitation reaches the singularity.}

In Section \ref{rw} we give a brief account of Zerilli's approach to (even type)
 gravitational perturbations of Schwarzschild spacetime, stressing the problems that
 arise in the negative mass case. We exhibit the unstable modes found in \cite{dg}, 
and introduce the new field $\hat { \Psi}$, which is smooth for smooth metric perturbations, 
no matter the sign of $M$, and obeys an equation free of  singularities
for any  $M$.
The main results of this paper are listed in a theorem, proved in Section \ref{proof}, from which 
 the negative mass Schwarzschild spacetime linear instability follows as a corollary.
In Section \ref{num} we illustrate, by means of numerical integrations of the
 linearized equations, how the unstable linear mode found in \cite{dg} for the
 negative mass Schwarzschild spacetime is excited by  perturbations with different initial data.
Section \ref{sum} summarizes our results.

\section{Scalar gravitational perturbations of the Schwarzschild spacetime}\label{rw}
  In the Regge-Wheeler gauge \cite{rw}, the scalar
perturbations for the angular mode $(\ell,m)$  are
described by  four functions $H_0(r,t)$, $H_1(r,t)$, $H_2(r,t)$ and
$K(r,t)$, in terms of which the perturbed metric takes the form,
\begin{eqnarray}
\label{RW1}
ds^2 & = & -\left(1-\frac{2M}{r}\right)\left(1- \e H_0 Y_{\ell,m}\right)dt^2 +
2  \; \e  H_1 Y_{\ell,m} dt dr+\left(1-\frac{2M}{r}\right)^{-1}\left(1+ \e H_2 Y_{\ell,m}\right)dr^2
\nonumber \\
& & +r^2\left(1+ \e  K\, Y_{\ell,m}\right) \left( d \theta^2 +  \sin^2(\theta)
d \phi^2 \right)
\end{eqnarray}
where  
$Y_{\ell,m}=  Y_{\ell,m}(\theta,\phi)$ are standard spherical
harmonics on the sphere.
The linearized Einstein equations for the
metric (\ref{RW1}), obtained by disregarding terms of order $\e^2$ or higher,
 imply $H_0(r,t)=H_2(r,t)$, and a set of coupled
differential equations for $H_1$, $H_2$ and $K$.

Of particular interest to us is the following unstable solution  found in \cite{dg}  {\em for the negative mass case}:
\begin{eqnarray}
\label{RW24a}
 K(t,r) & = & \frac{ (\lambda + 1)(r-2M)^{k }}{6M}\exp \left[\frac{k(t- r)}{2|M|}\right]   \\
 H_1(r,t) & = & -H_2(t,r) = -\frac{\lambda (\lambda+1)
   [2(\lambda+1) r-6M] \; r  (r-2M)^{k-1}}{36M^2}  \exp \left[\frac{k(t- r)}{2|M|}\right] \nonumber
\end{eqnarray}
where
\begin{equation} \label{k}
k =\frac{ (\ell-1) \ell (\ell+1) (\ell+2)} {6}  ,
\end{equation}
and
\begin{equation} \label{lam}
\lambda =  \frac{(\ell-1)(\ell+2)}{2} .
\end{equation}

The above solution has the following properties (see Section 7 in \cite{dg}): (i) it is exponentially growing in time, 
(ii) it is smooth for $r>0$,
 exponentially decaying for large $r$;
(iii) it has a fast decay as $r \to 0^+$ that guarantees that the first order algebraic and 
differential invariants of the Riemann
tensor do not diverge faster than their zeroth order piece, a condition 
of self consistence of the perturbation procedure, (iii)
it has finite gravitational energy $E_G$
\begin{equation} \label{eg}
E_G = - \frac{1}{8 \pi} \int_{\Sigma_{(3)}} G^{(2)}_{ab} \eta^a \zeta^b d
\Sigma_{(3)},\hspace{1cm}\eta = (1-2M/r)^{-1/2} \partial / \partial t,
\;\;\; \zeta = \partial / \partial t,
\end{equation}
where $G^{(2)}_{ab} $ is the second order correction to the Einstein
tensor, and $\Sigma_{(3)}$ the spacelike hypersurface orthogonal to $\eta^a$ (for details
see \cite{ghi,dg,chandra}.)

 As shown below, the evolution of generic initial data  with compact support away 
from the singularity will excite these singular modes, which implies that  the negative mass 
Schwarzschild spacetime
is linearly unstable. We first recall Zerilli's approach to the linearized problem, 
in order to exhibit
the difficulties in dealing with the evolution of initial data for the Schwarzschild spacetime 
in the negative mass case, and develop
an alternative approach to the linearized problem that allows us to overcome these problems.

\subsection{Solution of the even mode linearized Einstein equations: Zerilli's approach}
The linearized Einstein's equation for (\ref{RW1}) give a coupled system of partial differential equations
involving $K, H_1$ and $H_2$ \cite{rw}. This system
  can be  decoupled by introducing
the Zerilli function $\Psi_z(t,r)$, by the replacements \cite{zerilli},
\begin{eqnarray}
\label{RW2}
K & = & q(r) \Psi_z+\left(1-\frac{2M}{r}\right)\frac{\partial \Psi_z}{\partial r} \nonumber \\
H_1 & = & h(r)\frac{\partial \Psi_z}{\partial t}+r\frac{\partial^2 \Psi_z}{\partial t \partial r}
 \\
H_2 & = & \frac{\partial }{\partial r}\left[\left(1-\frac{2M}{r}\right)\left(h(r) \Psi_z
+r\frac{\partial \Psi}{ \partial r} \right)\right] -K  \nonumber
\end{eqnarray}
where $\lambda$ is defined in (\ref{lam}) and
\begin{eqnarray}
\label{RW2a}
q(r) & = &  \frac{\lambda(\lambda+1)r^2+3 \lambda M r + 6 M^2}{r^2(\lambda r + 3 M)}, \\
h(r) & = &  \frac{\lambda r^2 - 3 \lambda r M -3 M^2}{(r-2M)(\lambda r + 3M)}.\nonumber
\end{eqnarray}
Note that the  relations (\ref{RW2}) can be inverted and give
\begin{equation}
\label{zeri}
\Psi_z(r,t)  = \frac {r (r-2M)}{ ( \lambda+1 )  ( \lambda
\,r+3\,M ) } \left(H_2
 - r  \frac {\partial K}{\partial r}
\right)
    +\frac{r}{\lambda+1} K.
\end{equation}
The full set of linearized Einstein's equations then reduce to Zerilli's wave equation
\begin{equation}
\label{zerilli}
\frac{\partial^2 \Psi_z}{\partial t^2}+{\cal H} \Psi_z = 0,
\end{equation}
where,
\begin{equation} \label{scho}
{\cal H} = -\frac{\partial^2 }{\partial x^2} +V
\end{equation}
looks like a  quantum Hamiltonian operator  with potential
\begin{equation}
\label{poten1}
V = 2 \left(1 -\frac{2 M}{ r}\right) {\frac{ \lambda^2 r^2
\left[(\lambda+1) r + 3 M \right] + 9 M^2 (\lambda r +M)}{ r^3
(\lambda r + 3 M)^2} } ,
\end{equation}
and $x$ is the ``tortoise'' coordinate, related to $r$ by
\begin{equation}
\label{tor}
\frac{dx}{dr}=\left(1-\frac{2M}{r}\right)^{-1}.
\end{equation}
We will choose the integration constant such that $x=0$ at $r=0$, then
\begin{equation}\label{tortu}
x = r +2 M \ln \left| \frac{r-2M}{2M}\right|.
\end{equation}

\subsubsection{Case $M>0$, stability of the Schwarzschild black hole exterior metric}
For $M>0$,
the exterior static region
$r>2M$ of the Schwarzschild black hole gets mapped under (\ref{tortu}) onto $-\infty < x< \infty$, with
the black hole horizon sitting at $x=-\infty$. The potential $V$ in Zerilli's equation is positive definite and behaves as
$V \sim \exp(x/(2M))$ as $x \to -\infty$, $V \sim x^{-2}$ as $x \to \infty$ (see Figure 1). Equation (\ref{zeri})
indicates that a smooth metric perturbation with compact support in the exterior region corresponds to a smooth
Zerilli function in  $L^2({\mathbb R},dx)$. The fact that $\Psi_z^0:=\Psi_z \mid_{(t=0,x)}$ and $\dot \Psi_z^0 := \p/\p t \Psi_z
\mid_{(t=0,x)}$
 can be freely
chosen, together with (\ref{RW2}),  takes proper  account of the constraints among the initial data for $H_1,H_2$ and $K$.
To solve the Zerilli wave equation (\ref{zerilli}) from a given initial data
$(\Psi_z^0, \dot \Psi_z^0) \in L^2({\mathbb R},dx) \otimes L^2({\mathbb R},dx)$, we can use
that ${\cal H}$ is a self adjoint operator in $L^2({\mathbb R},dx)$
to expand $\Psi_z^0$ and $\dot \Psi_z^0$
 using  a complete set $\psi_E$ of eigenfunctions of ${\cal H}$ (${\cal H} \; \psi_E = E \psi_E$).  Equation (\ref{zerilli})
 then reduces
 to  the following ordinary differential equations for $a_E(t) := \int \psi_E(x) ^* \Psi_z(t,x) dx$:
\begin{eqnarray}\nonumber
 \ddot a_E &=& -E a_E \\  \label{a1}
 \dot a_E(0) &=& \dot a_E^0 := \int \psi_E ^*  \dot \Psi_z^0 \; dx  \\
 a_E(0) &=& a_E^0 := \int \psi_E ^* \Psi_z^0 \; dx .  \nonumber
\end{eqnarray}
whose solution is
\begin{equation} \label{a3}
a_E(t) = \begin{cases} a_E^0 \; \cos (\sqrt{E} t) + \dot a_E^0 \; E^{-1/2} \; \sin(\sqrt{E} t) \ & , E>0\\
a_E^0 + t \; \dot a_E^0 &, E=0 \\
a_E^0 \; \cosh (\sqrt{-E} t) + \dot a_E^0 \; (-E)^{-1/2}\; \sinh(\sqrt{-E} t) & , E<0
\end{cases}
\end{equation}
Since the Zerilli Hamiltonian ${\cal H} $ is positive definite, we can use  the above equations to
obtain and $L^2$ bound for $\Psi$ at time $t$ in terms of its data \cite{wald} as
\begin{equation} \label{bound}
\int |\Psi|^2 dx \leq 2 \left( \int  |\Psi^0|^2 dx + \int \overline{ \dot \Psi^0} \; {\cal H}^{-1} \dot \Psi^0 dx \right),
\end{equation}
where the inverse of ${\cal H}$  is defined using its spectral decomposition.
The detailed analysis in \cite{wald} gives also the following {\em uniform} bound for the Zerilli function
in terms of the initial data
\begin{equation} \label{wally}
|\Psi_z(t,x)|^2 \leq \int \left( |\Psi_z^0|^2 + \frac{1}{2} \overline{\Psi_z^0} \; {\cal H} \;  \Psi_z^0  +   \frac{1}{2} |\dot \Psi_z^0|^2
+\overline{ \dot \Psi^0} \; {\cal H}^{-1} \dot \Psi^0
\right) dx
\end{equation}
 This proves that the exterior, static region of a Schwarzschild black hole is stable.\\

\begin{figure}
\centerline{\includegraphics[height=8cm,angle=-90]{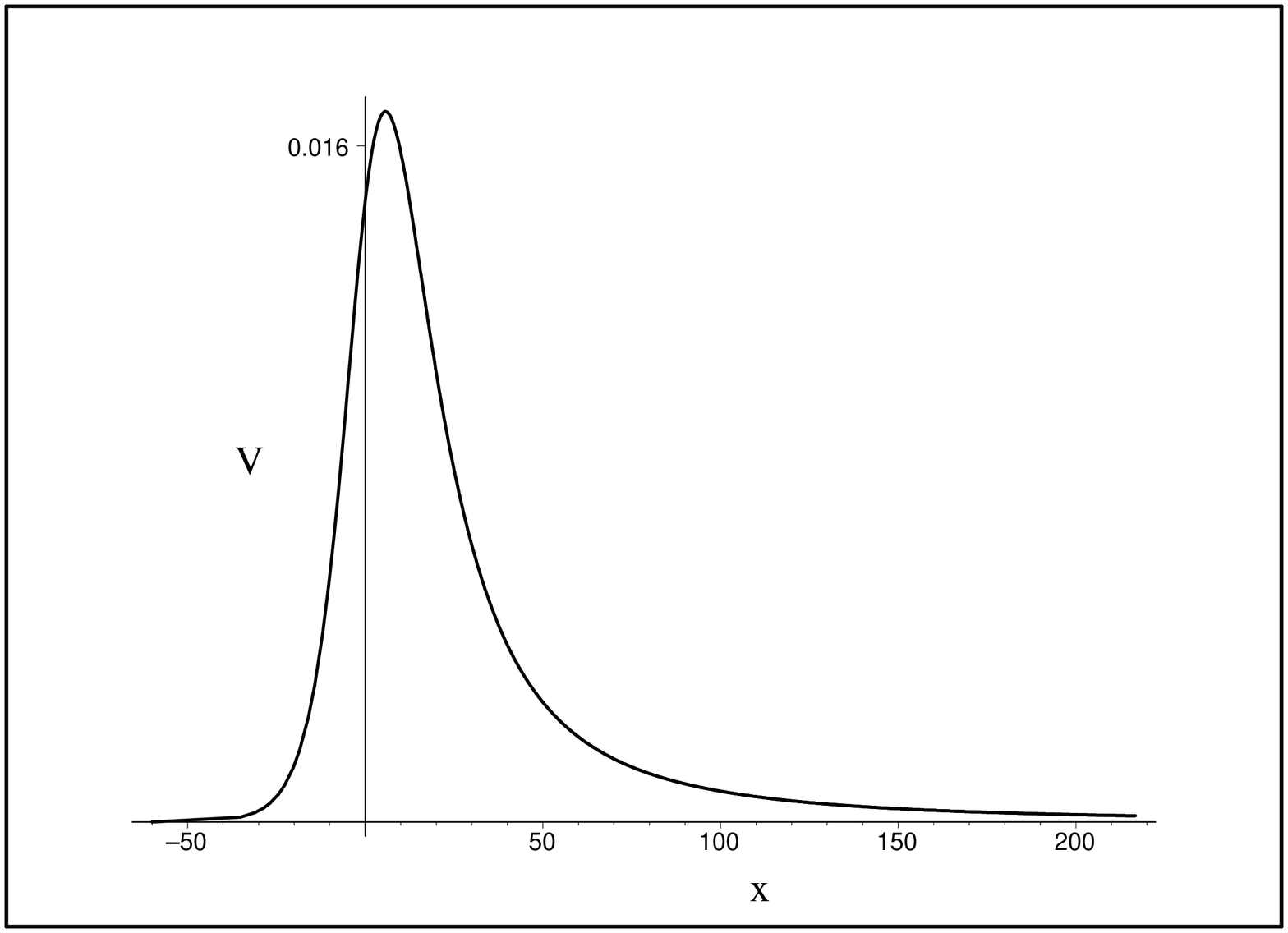} \hspace{1cm}
\includegraphics[height=8cm,angle=-90]{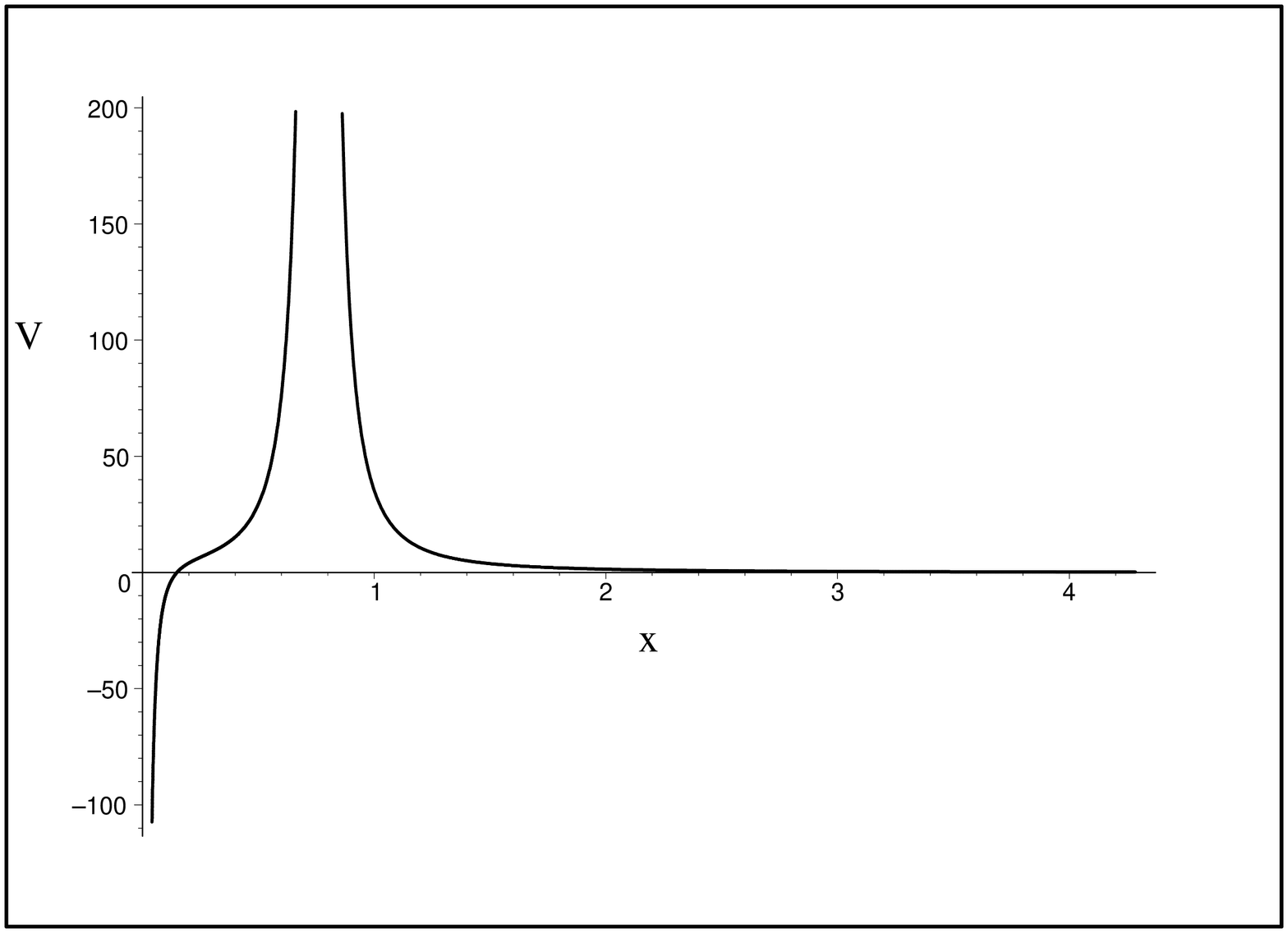}}
\caption{The left panel shows the  $\ell=2$ Zerilli potential for $M=3$ as a function of $x$. The black hole horizon is located at
 $x=-\infty$. Note that the potential is  smooth and positive. The right panel shows the  $\ell=2$ Zerilli potential for $M=-2$.
  The naked singularity is located at $x=0$. The ``kinematic'' double pole is at $x \simeq 0.761$.}
\end{figure}

\subsubsection{Case $M<0$, stability of the Schwarzschild naked singularity}

For $M<0$ the range of interest is $r>0$ (then $x>0$), and a number of difficulties arise due to the fact that
$q$ and $h$ in (\ref{RW2a}) are  singular at $r=r_s:=-3M/\lambda >0$, and that this point belongs
to the domain of interest. This implies that  $\Psi_z$ is singular at $r_s$, as
is also evident from equation (\ref{zeri}). The kind of singularity in $\Psi_z$ is characterized in
the following lemma:\\

\noindent
{\bf Lemma 1:} If $M<0$, a metric perturbation is smooth if and only if its Zerilli function at any fixed time
 is $C^{\infty}$ in open sets not containing $r_s$,
and admits a Laurent expansion
\begin{equation} \label{crc}
\Psi_z = \sum_{j \geq -1}  c_j (r-r_s)^j, \;\;\; c_0 = \frac{\lambda^2  c_{-1}}{3M (3 + 2 \lambda)}
\end{equation}
If initial data $\Psi_z^0$ and $\dot \Psi_z^0$ is given, both functions satisfying (\ref{crc}), the evolution
equation will preserve (\ref{crc}), i.e., this condition will hold at all times.\\

\noindent
{\bf Proof:} From (\ref{RW2}) and (\ref{RW2a}), the metric perturbation will be smooth
if and only if $h(r) \Psi_z
+r\frac{\partial \Psi_z}{ \partial r} $ and $q(r) \Psi_z+ (1-2M/r)\frac{\partial \Psi_z}{\partial r}$
are smooth. Both conditions lead to (\ref{crc}), added to smoothness in open sets not containing $r_s$.
A straightforward calculation shows that if $\psi$ satisfies (\ref{crc}) then so does ${\cal H} \psi$.
This guarantees that this condition will hold
at later times if it is satisfied by the initial data $\Box$\\

In particular, Zerilli functions for smooth metric perturbations are generically not square integrable  (no matter
which measure we use, either $dx=dr/(1-2M/r)$ or $dr$) due to the
pole in (\ref{crc}). As an example, the smooth $M<0$ metric perturbation (\ref{RW24a}) has the singular Zerilli function
\begin{equation}
\label{ines}
 \Psi_z^{unst}  = \frac{r \left( r - 2 M \right) ^{k}}{2 \lambda r +6M} \exp \left[\frac{k(t- r)}{2|M|}\right] =:
 \exp \left[\frac{kt}{2|M|}\right] \psi^{unst}.
\end{equation}

Given that $\Psi_z$ is a singular function of the metric perturbation, it is not a surprise
that the coefficients of the differential equation it obeys are singular. This explains
the second order pole of the Zerilli potential at $r_s$ (and the name ``kinematic'' given in \cite{dg}
to this singularity.) Note that the approach for solving Zerilli's equation in the $M>0$ case
completely breaks down when $M<0$ since (i) $\Psi_z \not \in L^2((0,\infty),dx)$ and
(ii) $V$ has the kinematic singularity. In particular, the associated quantum mechanical problem
with Hamiltonian ${\cal H}$ and domain $x \in (0,\infty)$ is not relevant in this case because of (i),
as discussed in detail in Section 7 of \cite{dg}.
Furthermore, since (\ref{zerilli})  is a wave equation in the half space $x>0$, we need to specify boundary conditions at
 $x=0$, besides the initial values of $\Psi_z^0$ and $\dot \Psi_z^0$,  to have  a unique solution.
 The fact that the potential has a singularity at the boundary,
 \begin{equation} \label{vcero}
 V \simeq -1/(4x^2)+ ... \;\; \text{ for } \; x \to 0^+.
 \end{equation}
implies that there is an infinite number of (formally, i.e., ignoring the kinematic singularity) self-adjoint extensions
of ${\cal H} := -\p^2/\p x^2 +V$, parameterized by $\theta \in S^1$, obtained  by demanding that the Zerilli function
behaves as
\begin{equation} \label{sae}
\Psi_z \simeq \cos (\theta) \left[ \left(\frac{x}{|M|}\right)^{1/2}  + ... \right] +\sin (\theta) \left[ \left(\frac{x}{|M|}\right)^{1/2}  \ln \left(\frac{x}{|M|} \right) + ... \right]
\end{equation}
for $x \gtrsim 0$ (the terms in square brackets are the leading terms of two linearly independent local solutions
of the eigenvalue equation ${\cal H} \Psi = E \Psi$, $E$ shows up at higher orders).
Note that both linearly independent local solutions in (\ref{sae}) are square integrable near $x=0$, the potential
belongs to the ``limit circle class'' at $x=0$  \cite{reed}.
This issue was
analyzed in detail in \cite{ghi} (see also  \cite{meetz} and \cite{reed})
 where it was concluded that $\theta=0$ is a physically motivated
choice, since it corresponds to finite energy perturbations with first order contributions
to the Kretschmann invariant not diverging faster than its zeroth order piece. These results
were confirmed in \cite{dg}, where it was further shown that every algebraic and some of the differential
invariants made out of the Riemann tensor share this property with the Kretschmann invariant. Given that this guarantees
the self consistency of the linearized treatment, we will be restrict our attention to the case
$\Psi_z \sim x^{1/2}$ from now on.

The question left open in \cite{dg} is how to evolve initial perturbation data in the $M<0$ case, since the ${\cal H}$ mode
expansion technique used for $M>0$ does not apply to the $M<0$ case.  In the next Section we introduce a field $\hat \Psi$ which is smooth
for smooth metric perturbation, and evolves according to a wave equation with a smooth potential for {\em any} sign of $M$,
thus providing a solution to the initial value problem in the negative mass case.

   \subsection{Solution of the even mode linearized Einstein equations: alternative approach}

As explained above, the quantum mechanical problem associated to ${\cal H}$
 is not directly relevant to the gravitational perturbation problem when $M<0$.
Zerilli's function
succeeds in reducing the full set of linearized Einstein's equations to a single wave equation, however, for $M<0$, this function
is singular in the relevant $r>0$ range. The ``kinematic'' singularity at $r_s=-3M/\lambda$ in (\ref{zeri}) indicates
that physically acceptable Zerilli functions have a simple pole at $r_s$ (Lemma 1).
Thus, even if  ${\cal H}$ could be extended to a self adjoint operator in some subspace of  $L^2({\mathbb R},dx)$,
this space would not be the natural setting for physically acceptable gravitational perturbations, which, because of the kinematic singularity,
 correspond, generically, to functions that are not square integrable.\\

In terms of the Zerilli function, the evolution problem in the negative mass case is: given data $\Psi_z^0, \dot \Psi_z^0$ both satisfying
the conditions in Lemma 1, and vanishing as $x^{1/2}$  when $x \to 0^+$ (i.e., (\ref{sae}) with $\theta=0$),
 find $\Psi_z$ for later times. The approach of solving this problem by separation of variables in Zerilli's
 wave equation and expanding by ${\cal H}$ modes
fails.
A satisfactory solution to the evolution problem requires
finding a new field $\hat \Psi$ that decouples the linearized equations obeyed by $K, H_1$ and $H_2$ -as $\Psi_z$ does-, and
that is a smooth function of $K, H_1$ and $H_2$ for {\em any} value of $M$.
In this section we show how this is done. We will state without proof our main result (Theorem  below), and illustrate
for the $\ell=2$ mode, case in  which the explicit formulae are relatively simple. We will defer the proof of the theorem
to section \ref{proof}\\

\noindent
{\bf Theorem:} Let $\psi_0$ be the solution of  ${\cal H} \psi_0 =0$ given  in equations (\ref{zm2}) and (\ref{zm3}).
 Define $g:=\psi_0{}'/\psi_0$
(a prime denotes derivative with respect to $x$) and the operators
\begin{eqnarray}\label{hp2}
\I  &:=& \frac{\p}{\p x} - g , \\
\hat \I  &:=& \frac{\p}{\p x} + g .
\end{eqnarray}
Let $V$ be the potential in Zerilli's equation. Then:
\renewcommand{\labelenumi}{(\roman{enumi})}
\begin{enumerate}
\item $$\I \; \left[ \frac{\p^2 }{\p t^2} - \frac{\p^2 }{\p x^2} + V \right]  =   \left[ \frac{\p^2 }{\p t^2} - \frac{\p^2 }{\p x^2}
+ \hat V \right] \; \I$$
with $\hat V = V - 2 g'$
smooth in the relevant domain ($r>2M$ if $M>0$, $r>0$ if $M<0$).
\item For any value of $M$, a  metric perturbation (\ref{RW1}) with Zerilli function $\Psi_z$
is smooth if and only if $\hat \Psi := \I \Psi_z$ is smooth in the relevant domain.
\item For $M<0$,
$\hat V \simeq 3/(4x^2)$ as $x \to 0^+$. As a consequence
 $\hat {\cal H}:=  \frac{\p^2 }{\p x^2} + \hat V(x)$
has a {\em unique} self adjoint extension in a domain ${\cal D} \subset L^2((0,\infty),dx)$, defined by the boundary condition
$\hat \Psi \simeq x^{3/2}$ as $x \to 0^+$ (see \cite{meetz},\cite{reed}).
 Moreover, for $\Psi_z$ as in (\ref{sae}), $\I \Psi_z \in {\cal D}$ if and
only if $\theta=0$. Thus ${\cal D}$ is the set of  physically relevant perturbation functions $\hat \Psi$.
\item Assume that $M<0$ and  that $(\Psi_z^0, \dot \Psi_z^0)$ is an appropriate initial data set, i.e., it satisfies
 the conditions in Lemma 1 and
the boundary condition $\theta=0$ in (\ref{sae}). Note from (iii) that both $\I \Psi_z^0$ and $\I \dot \Psi_z^0$ belong to
${\cal D}$.  Let
 $\hat \Psi$ be the unique solution in ${\cal D}$
for the wave equation $\left[ \frac{\p^2 }{\p t^2} + \hat {\cal H} \right] \hat \Psi =0$ on the half space $x>0$,
subject to the initial conditions $\hat \Psi \mid_{(t=0,x)}= \I \Psi_z^0$ and $\p/\p t \hat \Psi
\mid_{(t=0,x)} = \I \dot \Psi_z^0$.
This solution can be obtained by $\hat {\cal H}$ mode expansion as is done in   equations (\ref{a1}) and (\ref{a3}).
 The Zerilli field at all times is then given by
\begin{equation}  \label{conv}
\Psi_z(t,x) = \int_0^t \left( \int_0^{t'} \hat \I \hat \Psi(t'',x) dt'' \right) dt' + t \dot \Psi_z^0 + \Psi_z^0.
\end{equation}
\end{enumerate}

Let us clarify some aspects related to the above theorem. Generically,
$\Psi_z$ has a pole at $r_s$ (Lemma 1), and so does $g$, then the operators $\I$ and $\hat \I$
are singular. The singularities cancel in such a way  that $\hat \Psi := \I \Psi_z$ is smooth in the
domain of interest, that is,
$\I$ removes the singularity in $\Psi_z$. In the same way $-2g$ subtracts the pole in $V$
to produce a smooth $\hat V$. As an example, for $\ell=2$ we have \cite{ghi}
\begin{equation} \label{l2zm}
\psi_0 = \frac{r(r^3+3Mr^2-6M^3)}{8M^4(3M+2r)}
\end{equation}
and thus
\begin{equation}
\hat V = \frac{6 (r-2M)(2r^7+5Mr^6-9M^2r^5-33M^3r^4-24M^4r^3+36M^5r^2+36M^6 r-36M^7)}{r^4 (r^3+3Mr^2-6M^3)^2},
\end{equation}
whose only singular point  lies at $r \simeq 1.2 M$, which is outside the domain of interest both for
positive or negative $M$. The $\ell=2$ 
 unstable mode (\ref{ines}) for $M<0$ becomes
\begin{equation}
\label{hpsiunst}
\hat \Psi ^{unst}  = e^{2(t-r)/|M|} \frac{r^ 3 \left( r-2M \right)^4}{4r^3+12M r^2+24M^3} =: e^{2t/|M|} \hat \psi^{unst},
\end{equation}
which is also $C^{\infty}$ in the domain of interest. In Figure 2  we exhibit $\hat V$ for $\ell=2$
and $M=3$ (left), and $\hat V$ for $\ell=2$ and $M=-2$,   superposed with the unstable mode $\hat \psi^{unst}$ (right).\\
 Since the Einstein's linearized equations  reduce to the single equation
 \begin{equation} \label{nee}
\left[ \frac{\p^2 }{\p t^2} + \hat {\cal H} \right] \hat \Psi =0, \;\;\; \hat \Psi \in  {\cal D}
\end{equation}
and
 $\hat {\cal H}$ is self adjoint in ${\cal D}$, we can then solve this equation
by $\hat {\cal H}-$mode expansion, in the same way as is done with $\Psi_z$ when $M>0$. 
This gives an answer to the issue of evolution in  the non globally hyperbolic spacetime 
in a way entirely analogous to that developed in \cite{wi}, the only difference being that 
the radial part of the equations dealt with in \cite{wi} are {\em positive}, essentially self
adjoint operators. 
Note that we can work entirely in ``$\hat \Psi$-space'' with no reference to
the Zerilli function (as will be done in the next section), and that our alternative formulation also works in the positive mass case.
The  usefulness of equation (\ref{conv}) lies in the simpler connection that there is between the Zerilli field and the
perturbed metric elements $H_1,H_2$ and $K$, equations (\ref{RW2}). If we want to construct the perturbed metric elements from
$\hat \Psi$, the shortest way seems to be inserting (\ref{conv}) in (\ref{RW2}). Note that (\ref{conv}) is {\em not}
 an evolution equation. It just tells us how to recover the information lost  after applying $\hat \I \I$ to
 the Zerilli function (see
 Section \ref{proof}).  In fact, we need to solve first the evolution problem for $\hat \Psi(t,x)$, and then use 
the solution $\hat \Psi(t,x)$  in the integrand in (\ref{conv}) to obtain the corresponding solution for $\Psi_z(t,x)$.\\

\noindent
{\bf Corollary:} The negative mass Schwarzschild solution is unstable.\\

\noindent {\bf Proof:} The spectrum of the operator $\hat {\cal H}$ in ${\cal D}$ contains the negative eigenvalue $-k^2$
($k$ given in (\ref{k})), with eigenvector $\hat \psi ^{unst} = C \; \I \psi^{unst}$, where $\psi^{unst}$ is
given in (\ref{ines}) and  $C$ is a normalization constant.
Let $a_{-k^2}^0$ and  $\dot a_{-k^2}^0$ be the projections of $\hat \Psi_0$ and $\dot {\hat{\Psi}}_0$  onto this mode, (see equation (\ref{a1})), then from (\ref{a1}) and  (\ref{a3}) applied to (\ref{nee}) we obtain
\begin{equation} \label{decomp}
\hat \Psi (t,x)  = \left[ a_{-k^2}^0 \; \cosh (k t) + \frac{\dot a_{-k^2}^0}{k} \; \sinh(k t) \right] \hat \psi^{unst}(x)
+ \Phi(t,x),
\end{equation}
where $\Phi(t,x)$ is a linear combination of modes $\psi_E$, with $E \neq  -k^2$. Since the above decomposition is orthogonal
$$\int_0^{\infty}  |\hat \Psi |^2 dx > \left[ a_{-k^2}^0 \; \cosh (k t) + \frac{\dot a_{-k^2}^0}{k} \; \sinh(k t) \right]^2$$
and thus exponentially growing for large $t$ $\Box$.\\

Numerical evidence indicates that $\hat \psi^{unst}$ is the only negative  eigenvalue
  of $\hat {\cal H}$. If this is the case, then $\Phi$ is bounded and uniformly bounded
in a similar way as the positive mass Zerilli function is, equations  (\ref{bound}) and (\ref{wally}).

\begin{figure}[h,t]
\centerline{\includegraphics[height=8cm,angle=-90]{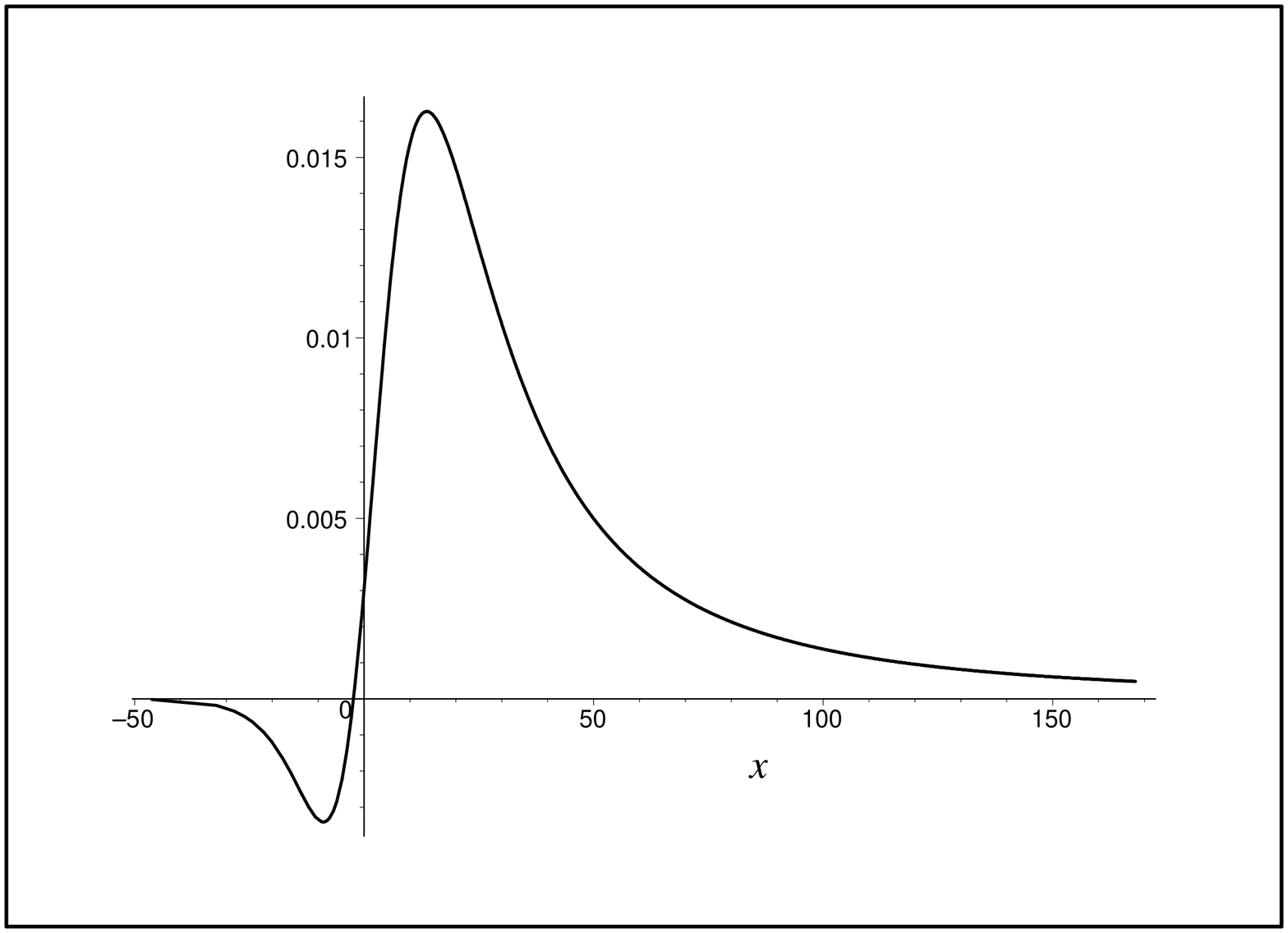} \hspace{1cm}
\includegraphics[height=8cm,angle=-90]{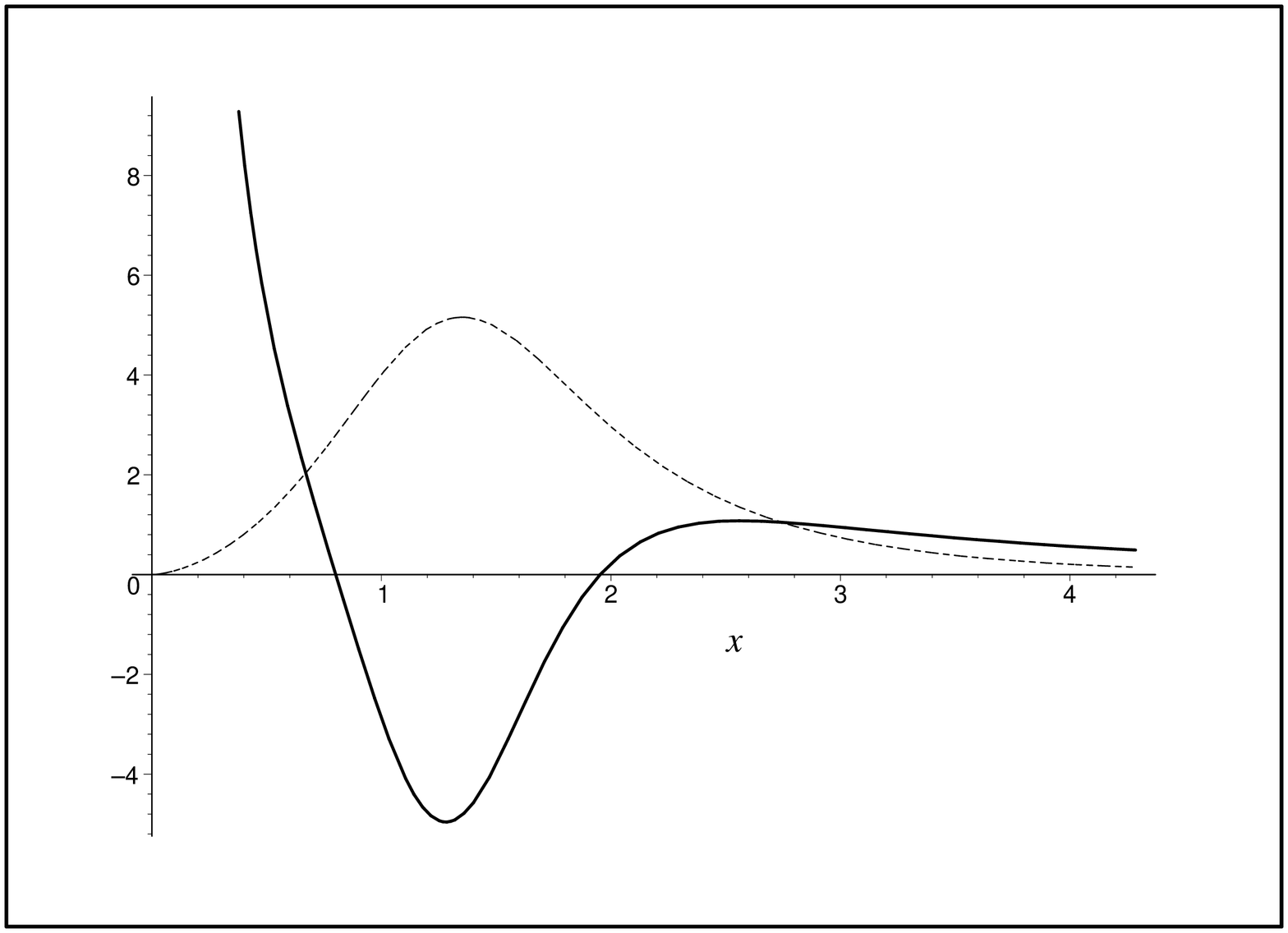}}
\caption{The left panel shows the potential $\hat V$, for $\ell=2$ and positive mass $M=3$, as a function of $x$.
 The black hole horizon is located at
 $x=-\infty$. The right panel shows
 $\hat V$ for $\ell =2$ and $M=-2$ (solid line), and the unstable mode $\hat \psi_{unst}$ 
given  (\ref{hpsiunst})(dotted line).
This mode satisfies $\hat {\cal H} \hat \psi_{unst} = -\hat \psi_{unst}$. }
\end{figure}

\section{Numerical integration of the evolution equations} \label{num}

Numerical integrations of the wave equation
\begin{equation}
\left[\frac{\p^2}{\p t^2}-\frac{\p^2}{\p x^2} -\hat V \right] \hat \Psi = 0
\end{equation}
subject to the boundary condition $\hat \Psi \simeq x^{3/2}$ as $ x \to 0^+$ were carried out
for $\ell=2$, $M=-1$ using the Maple built in integrator for partial differential equations,
working in the standard radial coordinate.
The boundary  condition at $r=0$ was enforced by imposing Robin type  boundary conditions
in the form $3 \hat \Psi - \p \hat \Psi / \p r =0$ at $r = 10^{-4}$.
We  also set $\hat{\Psi} = 0$ at $r=10$
and restricted the initial data and evolution time so that this condition is trivially satisfied. In all cases we set $\dot{\hat{\Psi}}^0=0$ for simplicity. We evolved
different $\hat{\Psi}^0$ initial data sets to see how
the unstable (\ref{ines}) mode gets excited, and the resulting numerical solution $\hat \Psi$ was contrasted with
its expected projection $ a_{-k^2}^0 \; \cosh (k t) \; \hat  \psi^{unst}(x)$ onto the unstable mode.

In the case of
Figure 3,  $\hat \Psi^0 =  \exp(-10(r-2)^2) \Theta(r-0.0002)$, $\Theta$ a step function. If normalized,  this function
 gives a projection $a_{-k^2}^0 \simeq 0.79$ onto the unstable mode.
\begin{figure}[h] \label{evolu1}
\centerline{\includegraphics[height=8cm,angle=-90]{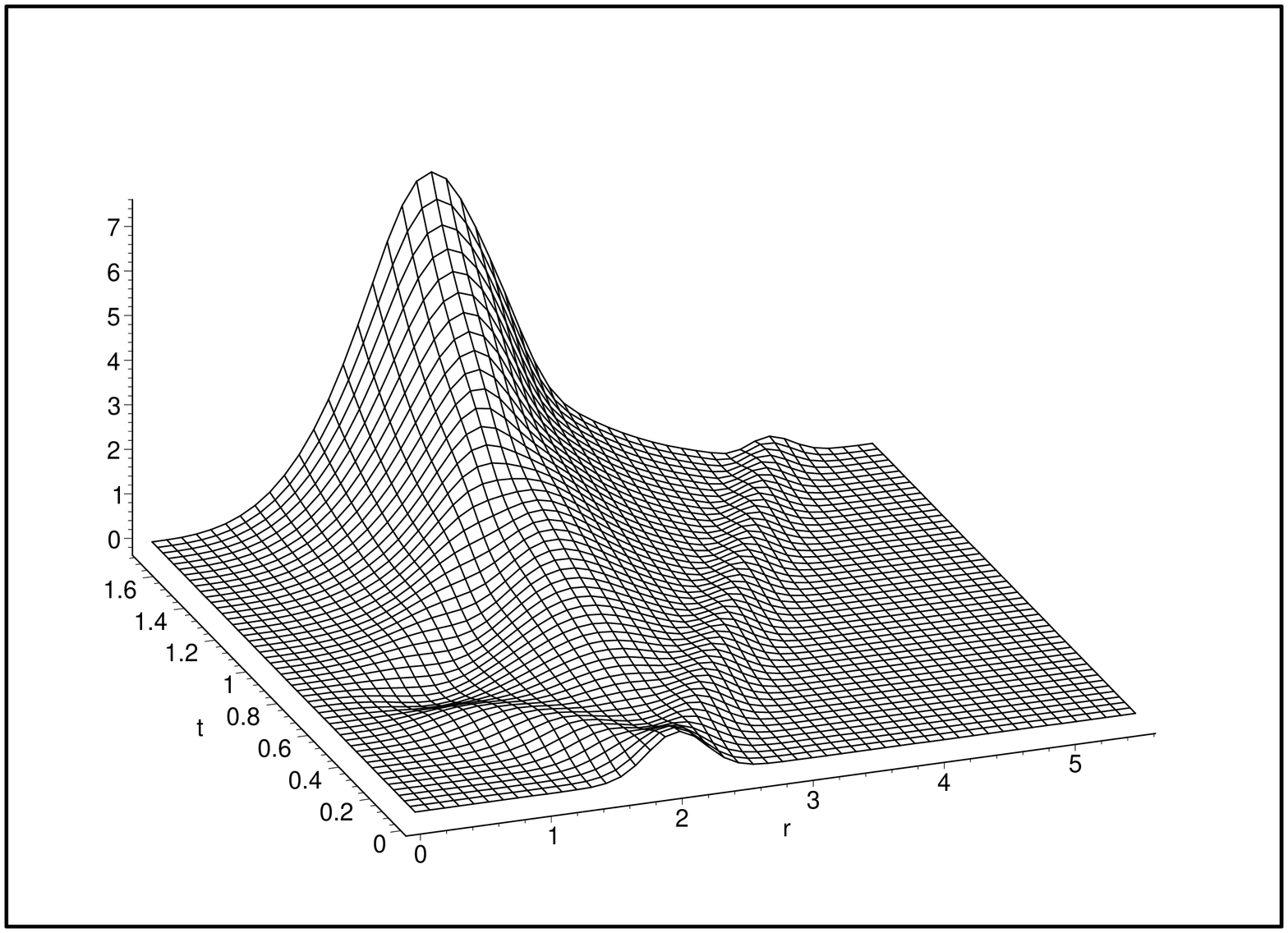} \hspace{1cm}
\includegraphics[height=8cm,angle=-90]{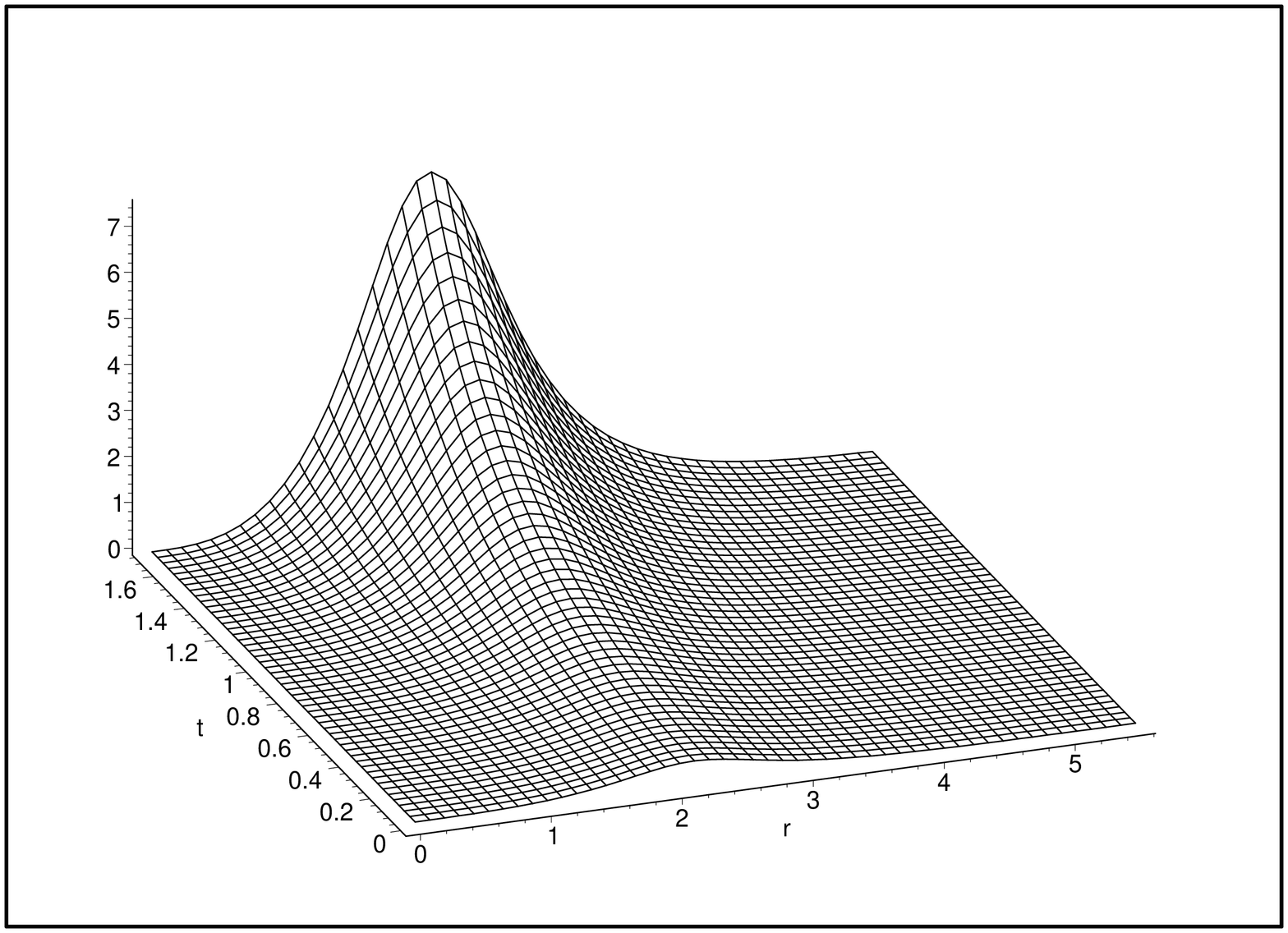}}
\caption{Left: evolution of initial data centered at $r =2$, with a strong projection onto the unstable mode. Right:
time evolved unstable mode component for this data.}
\end{figure}
The unstable mode dominates for $t \gtrsim 1.6$, and it is noticeable from $t=0$. The evolution of two stable wave packets
 moving oppositely  is also evident in the plot.

The left panel in Figure 4 shows the evolution  up to $t=2.7$ of the data 
 $\hat \Psi^0 =  \exp(-10(r-4)^2) \Theta(r-0.0002), \dot{\hat{\Psi}}^0=0$,
 which has a milder projection onto the unstable mode
($a_{-k^2}^0 \simeq 0.07$ when normalized). The right panel 
 contrasts $\hat \Psi(t=0,r), \hat \Psi(t=3,r)$ and
 the unstable mode properly scaled by the $\cosh(6)$ factor. Note that the unstable mode is  noticeable starting at $t \simeq  1.5.$

\begin{figure}[h] \label{evol2}
\centerline{\includegraphics[height=8cm,angle=-90]{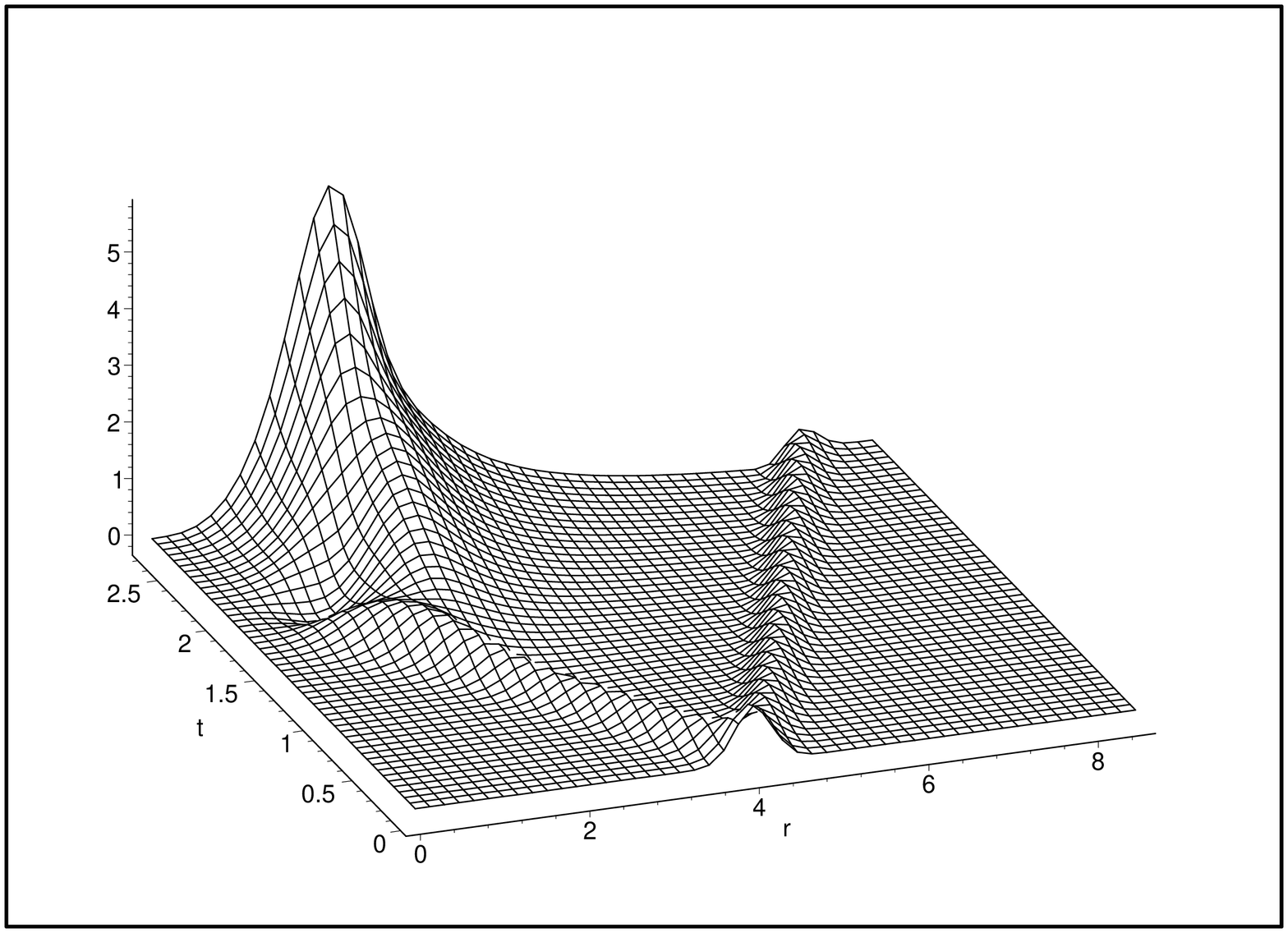} \hspace{1cm}
\includegraphics[height=8cm,angle=-90]{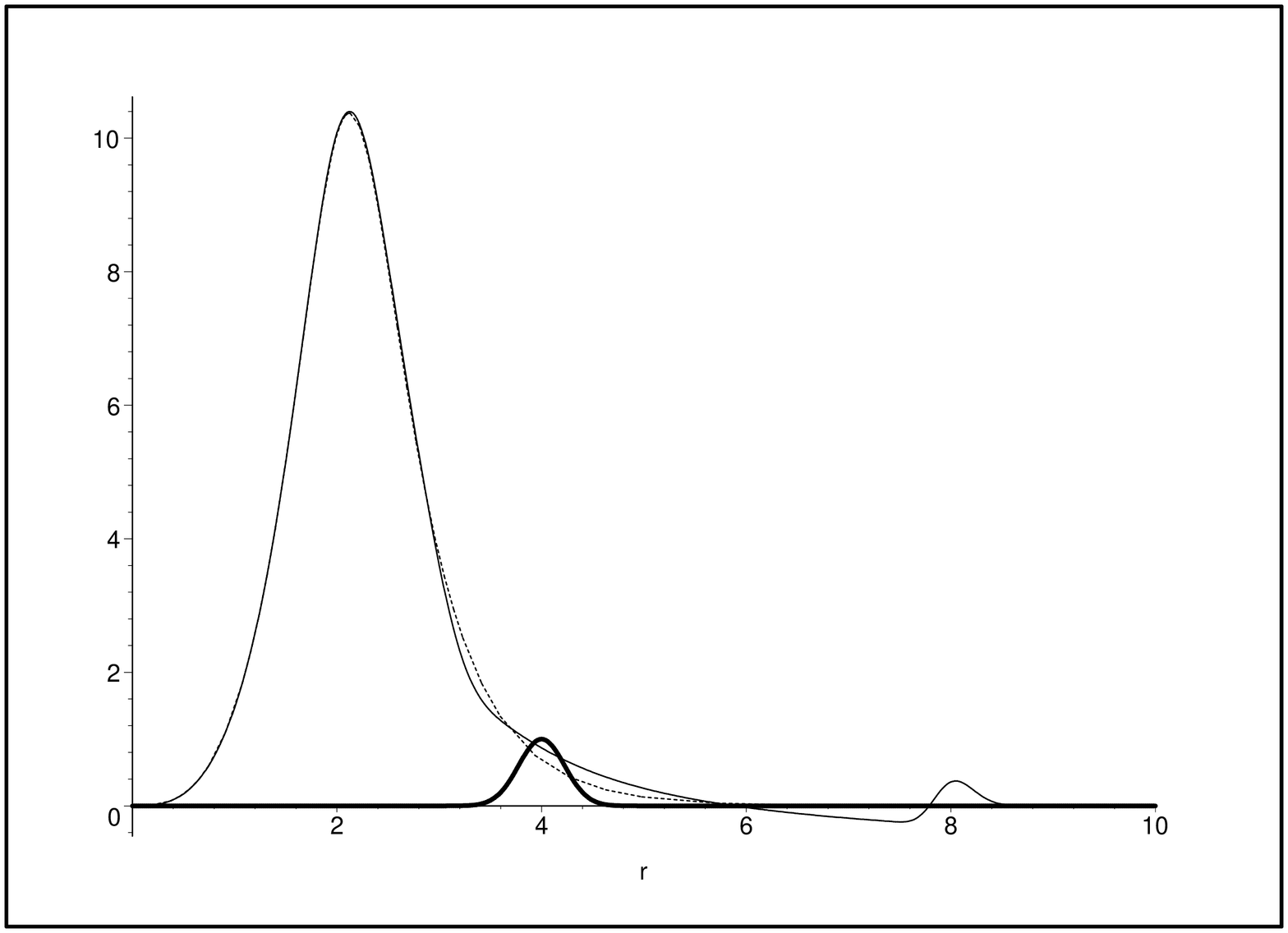}}
\caption{Left: evolution of initial data centered at $r =4$. Right:
$\hat \Psi$ at $t=0$ (thick solid line), and $t=3$ (thin solid line), contrasted to the
evolution of the unstable component at $t=3$ (dashed line).}
\end{figure}

To have a smaller overlap with the unstable mode we use $\hat \Psi(t=0,r) = -\exp(-10(r-4)^2)\; \Theta(r-0.0002) \; \sin(10(r-4))$
(Figure 5, left panel). The graph may mistakenly (see equation (\ref{decomp})) suggest that the unstable mode is not excited before
the ingoing wave packet reaches the singularity. The right panel in this figure  exhibits the evolution of the projection of the
initial data onto the unstable mode. This mode is initially highly suppressed because $a_{-k^2}^0 \simeq 6 \times 10^{-3}$

\begin{figure}[h] \label{evol3}
\centerline{\includegraphics[height=8cm,angle=-90]{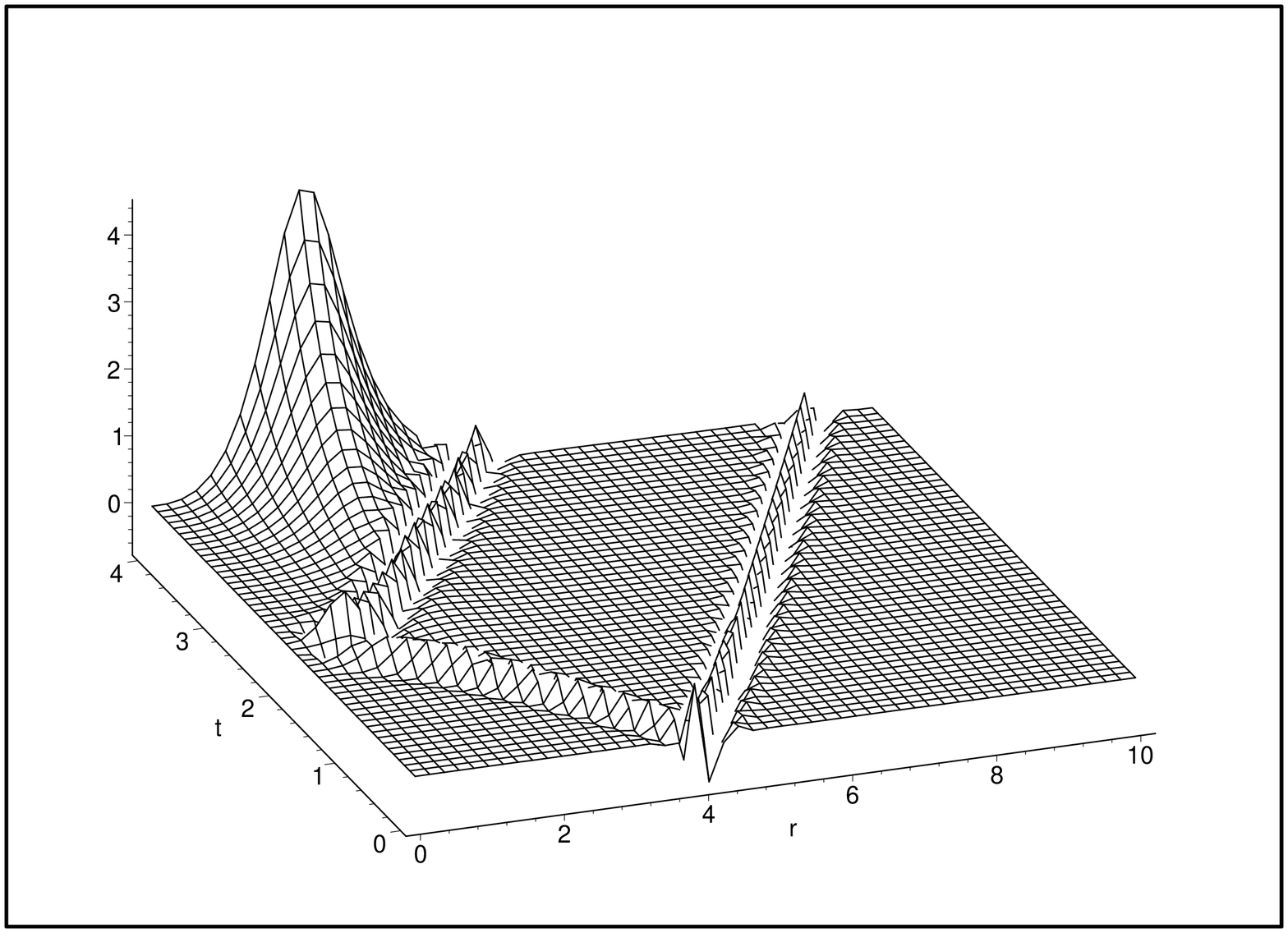} \hspace{1cm}
\includegraphics[height=8cm,angle=-90]{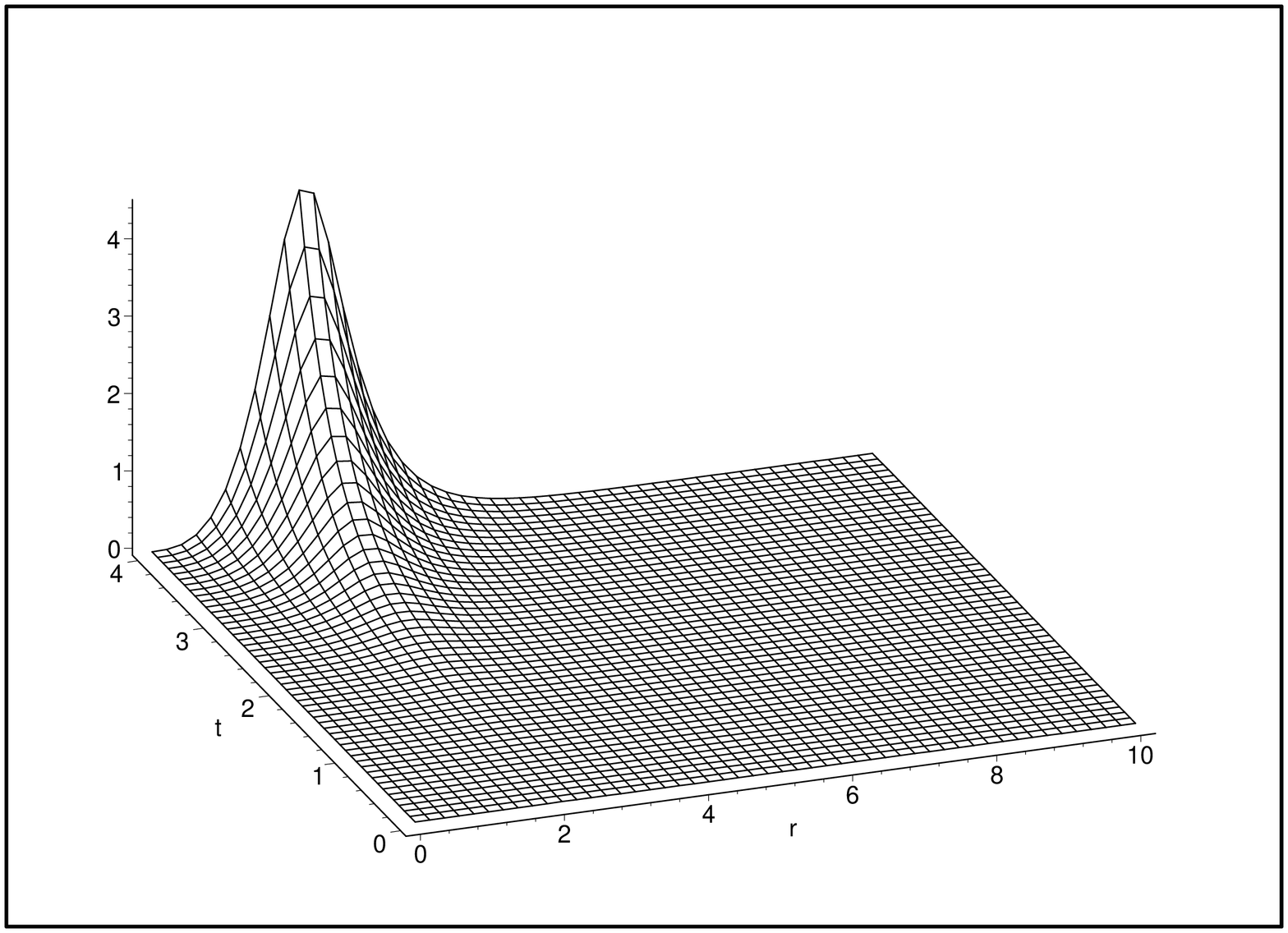}}
\caption{Left: evolution of initial data centered at $r =4$ and modulated with a sine function. The initial data has a
  weak projection onto the unstable mode. Right:
time evolved unstable mode component for this data.}
\end{figure}

Finally, the left panel in Figure 6 exhibits the evolution of data which is almost orthogonal ($a_{-k^2}^0 \simeq -2 \times 10^{-10},
\dot a_{-k^2}^0=0$) to the unstable mode.
The excitation of the unstable mode reaches an amplitude $\sim 2 \times 10^{-8}$ and so is unnoticeable in the displayed time range.
The right panel of the figure shows the initial data (dotted line), the result of the evolution at $t=1$ (thin solid line)
and the evolved data at $t=3$ (thick solid line).

\begin{figure}[h] \label{evol4}
\centerline{\includegraphics[height=8cm,angle=-90]{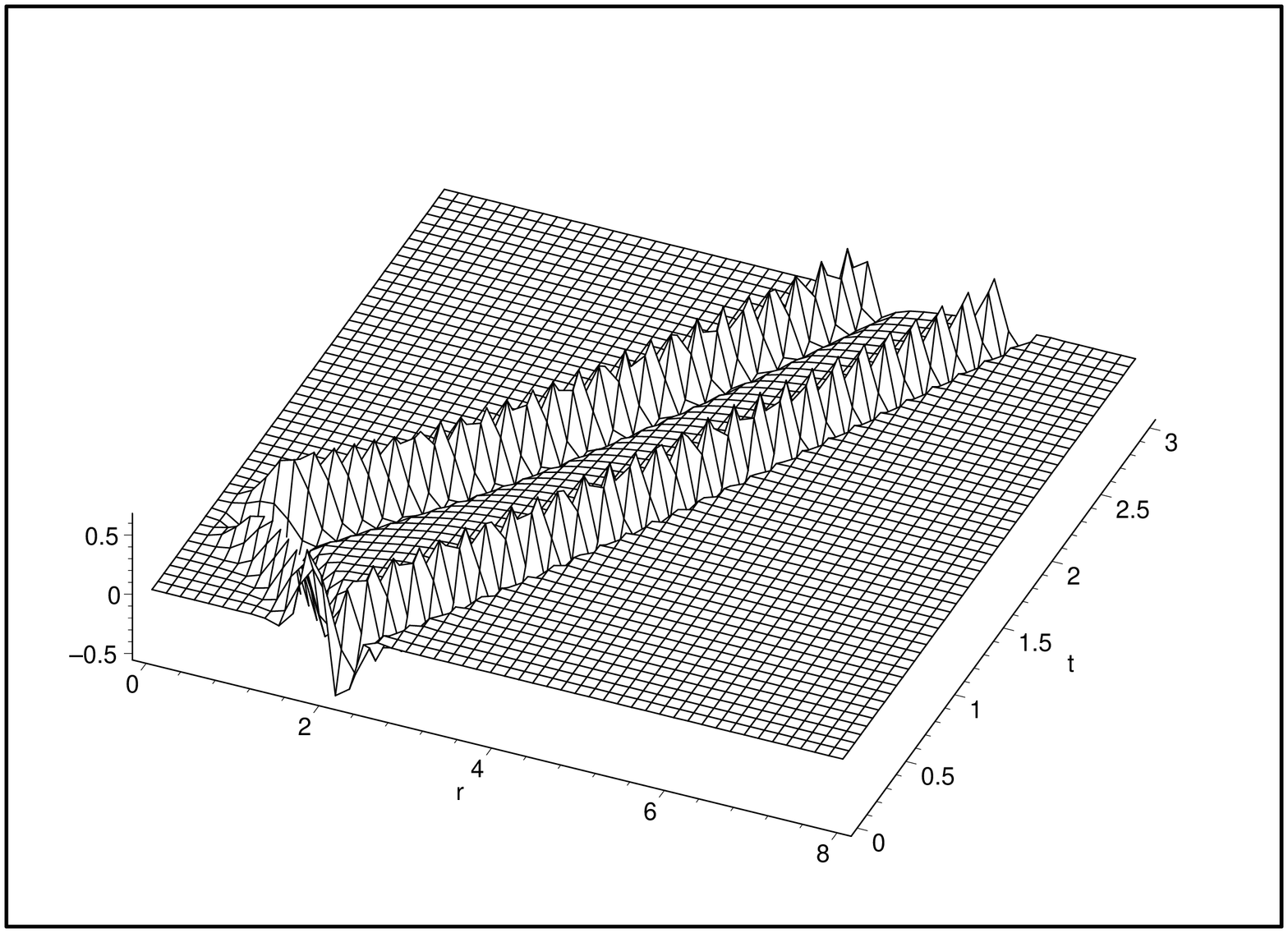} \hspace{1cm}
\includegraphics[height=8cm,angle=-90]{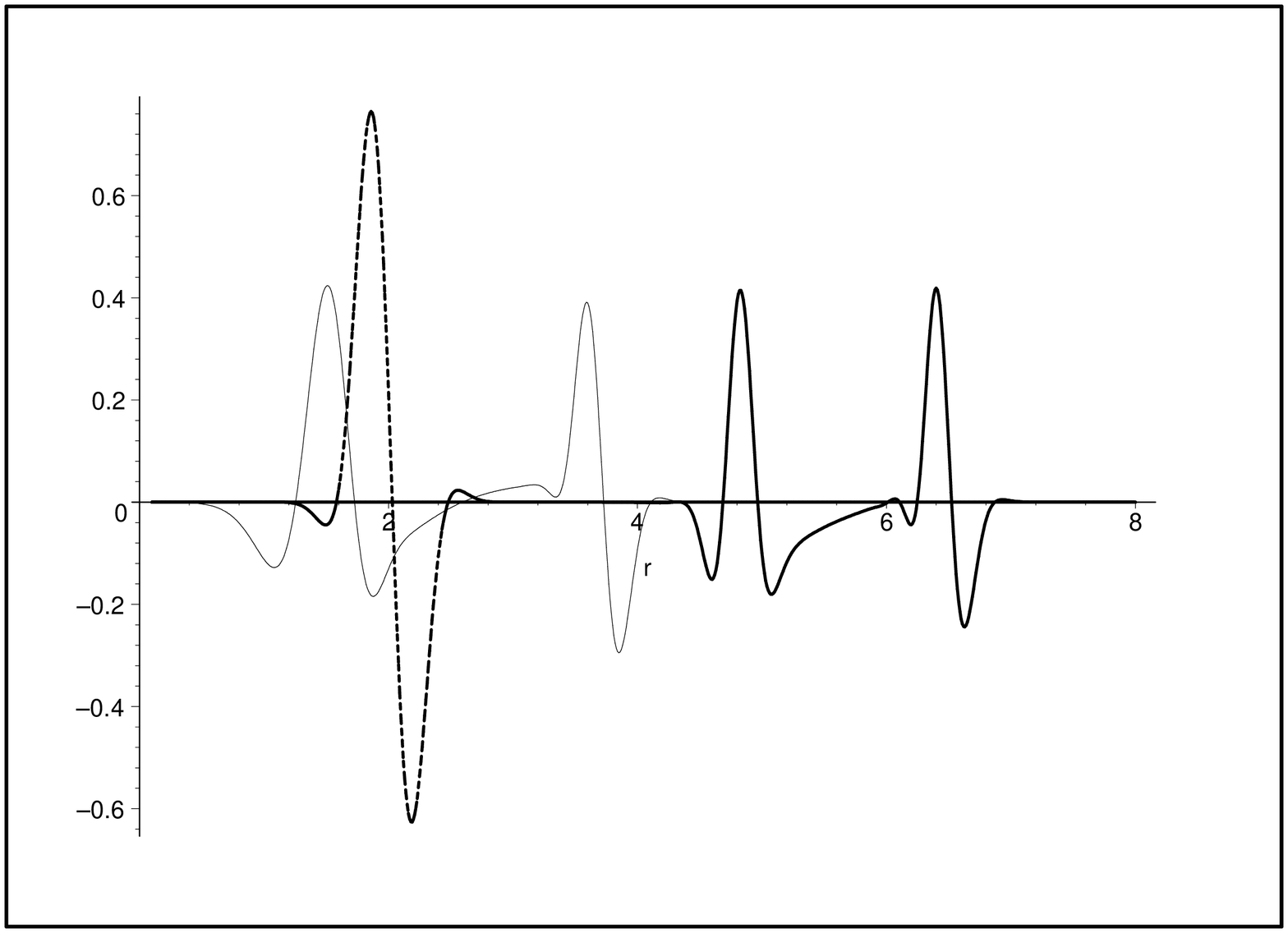}}
\caption{Left: evolution of initial data centered at $r =2$, modulated with a sine function and having a negligible projection onto the unstable mode.
Right: $\hat \Psi$ at $t=0$ (dotted line), $t=1$ (thin solid line) and $t=3$ (thick solid line).}
\end{figure}

\section{Proof of the Theorem} \label{proof}

The alternative $\hat \Psi$ to the Zerilli field as a means of  describing the even modes of linear gravitational perturbations
is suggested by  the ``intertwining'' potential technique in quantum mechanics \cite{price}, whose original motivation is
that of replacing a one dimensional quantum mechanical Hamiltonian with another Hamiltonian having a more elementary
potential. We have actually used an intertwiner to replace a potential {\em with a singularity at} $r_s$ with one
free of such singularity, with the added benefit that the resulting Hamiltonian has a unique self adjoint extension
that happens to agree with the boundary condition at  $r=0$  that is natural to the problem! Intertwiners appear
also the context of supersymmetric quantum mechanics \cite{cooper}, pairs of supersymmetric Hamiltonians
being related by an intertwiner constructed using a zero energy wave function.

\subsection{Intertwining operators} \label{inter}

Consider a two dimensional wave  equation with a space dependent potential $V$
\begin{equation} \label{z1}
\left[ \frac{\p^2 }{\p t^2} - \frac{\p^2 }{\p x^2} + V(x) \right] \Psi = 0,
\end{equation}
and a linear operator $\I= \frac{\p}{\p x} -g(x)$ such that \cite{price}
\begin{equation} \label{io}
\I \; \left[ - \frac{\p^2 }{\p x^2} + V(x) \right]  =   \left[ - \frac{\p^2 }{\p x^2} + \hat V(x) \right] \; \I
\end{equation}
for some potential $\hat V(x)$. Since $\I$  commutes with $\p /\p t$, any solution $\Psi$ of (\ref{z1})  gives a -possibly trivial- solution $\hat \Psi := \I \Psi$ for the equation
\begin{equation} \label{z2}
\left[ \frac{\p^2 }{\p t^2} - \frac{\p^2 }{\p x^2} + \hat V(x)
\right] \hat \Psi = 0.
\end{equation}
Separation of variables $\Psi = \exp(i\w t) \psi(x)$
  ( $\hat \Psi = \exp(i\w t) \hat \psi(x)$) reduces (\ref{z1})
and (\ref{z2})  to   Schr\"odinger like
equations
\begin{eqnarray} \label{z3}
{\cal H} \psi  &=& \left[ - \frac{\p^2 }{\p x^2} + V(x) \right] \psi = \w ^2 \psi ,\\
\hat {\cal H} \hat \psi &=& \left[ - \frac{\p^2 }{\p x^2} + \hat
V(x) \right] \hat \psi = \w ^2 \hat \psi , \label{z4}
\end{eqnarray}
If we do not specify boundary conditions, there will be
two linearly independent solutions of (\ref{z3}) for any chosen
complex $\w$. Let us   denote any two such solutions  as $\psi_{\w}^{(j)}, j=1,2$. Note from (\ref{io}) that
 $ \I \psi _{\w}^{(j)}$ are (possibly trivial) solutions of  (\ref{z4}).\\
The conditions for the existence of an intertwining operator can be
obtained
 by applying (\ref{io}) to an
arbitrary function $\psi$, and then isolating  terms in $\psi$ and $\psi' := \p \psi / \p x$ (the higher derivative terms
cancel out).
The coefficient of $\psi'$ gives
\begin{equation} \label{vh}
\hat V = V - 2 g' .
\end{equation}
Adding the condition from the  $\psi$ coefficient gives $(g' + g^2-V)' = 0$, i.e.
$g'+g^2=V-\w_o{}^2$ for some constant $\w_o$. This last condition is
more transparent if written in terms of  $\po :=\exp (\int^x g(x') dx' )$, which satisfies $\po'/\po=g$ and
\begin{equation} \label{zz}
\left[ - \frac{\p^2 }{\p x^2} + V \right] \po = \w_o ^2 \po .
\end{equation}
From this follows \cite{price},

\vspace{.3cm}

\noindent
{\bf Lemma 2:} From any solution of (\ref{zz}) it is possible to construct an intertwining operator
 $\I= \frac{\p}{\p x} -g(x)$  by choosing
$g = \po' / \po$. This gives  $\hat V = V - 2 g'$ in (\ref{io}).

\vspace{.3cm}

Lemma 2 collects the results we  need from \cite{price},
but  we need to elaborate further on these results to get some information  about
the possible ways to invert the effect of $\I$.  To fix the notation, let  $\psi_{\w_o}^{(j=1)}=\po$, and
$\psi_{\w_o}^{(j=2)}$ be a  linearly independent solutions of (\ref{zz}). The
kernel of $\I$ is the span of $\pe$, since   $0 = \I \psi = \psi' - \pe '
/ \pe \psi$ implies that $\psi$ is proportional to $\pe$. The form
of an intertwiner $\hat \I = \frac{\p}{\p x} -h(x)$
satisfying
\begin{equation} \label{iio}
\hat \I \; \left[ - \frac{\p^2 }{\p x^2} + \hat  V(x) \right]  =
\left[ - \frac{\p^2 }{\p x^2} +  V(x) \right] \; \hat \I
\end{equation}
can be guessed from Lemma 2 by noting that, since $\hat V - 2h' = V =
\hat V + 2g'$, the only possible way back to $V$ is that
${\cal H} (1 / \pe) \propto (1 / \pe)$. That this is actually the
case can be checked by a direct calculation
 using our previous results, from where we obtain
  $\hat {\cal H}  (1/ \pe) = \w_o{}^ 2 /\pe$. We will set $\hat \psi {}_{\w_o}^{(j=2)} := 1/\pe$ and
  choose $\hat \psi {}_{\w_o}^{(j=1)}$ such that $\hat I \hat \psi {}_{\w_o}^{(j=1)} = \pe$.
  It follows that $\hat \I := \frac{\p}{\p x} + g(x)$ satisfies (\ref{iio}), and a simple calculation shows that
$\hat \I \I \psi = (\w_o{}^2 - {\cal H}) \psi$, i.e., the non trivial kernels of $\I$ and $\hat \I$ combine
in such a way that the kernel of $\hat \I \I$ is the two dimensional $\w_o^2$ eigenspace of ${\cal H}$. Note that we have
shown that we can label the solutions of (\ref{zz}) and its hat version as
 such that
\begin{eqnarray} \nonumber
\pe &=&  \hat I \hat \psi _{\w_o}^{(j=1)}\\ \label{qsy}
I \psi _{\w_o}^{(j=2)} &=&  \hat \psi _{\w_o}^{(j=2)} = 1 / \pe  \label{pp}\\
I \psi _{\w_o}^{(j=1)} &=& \hat I \hat \psi _{\w_o}^{(j=2)} =0\nonumber
\end{eqnarray}

We have proved the following
\vspace{.3cm}

\noindent
{\bf Lemma 3:}  The kernel of $\I = \frac{\p}{\p x} - \psi _{\w_o}^{(j=1)} {}' / \psi _{\w_o}^{(j=1)}$
is the subspace spanned by $\po ^{(j=1)}$.  If $ \hat \I  =  \p/\p x +
 \psi _{\w_o}^{(j=1)} {}' / \psi _{\w_o}^{(j=1)}$, then (\ref{iio}) holds, also
 \begin{equation} \label{inverse}
 \hat \I \I = (\w_o^2 - {\cal H}),
\end{equation}
  and  the solutions of ${\cal H} \psi_{\w} = \w^2 \psi_{\w}$ and $\hat {\cal H} \hat \psi_{\w} = \w^2 \hat \psi_{\w}$
can be labeled such that the equations (\ref{qsy}) hold.\\

In the supersymmetric quantum mechanics context, $\w_o=0$ and $\psi _{\w_o}^{(j=1)}$ satisfies appropriate boundary conditions to make it
 an eigenfunction of ${\cal H}$. Moreover, it  corresponds to the lowest eigenvalue of ${\cal H}$.
In this case  ${\cal H}$ and $\hat {\cal H}$ are isospectral, except for $\w_o{}^2=0$, which is missing in the spectrum of $\hat {\cal H}$.
 Equations (\ref{pp}) then leads to the situation
depicted in Fig 2.1 in \cite{cooper}. In the above construction, however, we do not require any specific boundary condition
on the function $\pe$ used to construct the intertwining operator.

\vspace{.3cm}

The intertwining  operator (\ref{io}) will be useful whenever  $\hat V$ is simpler than
$V$. However, information is lost when solving (\ref{z2}) instead of (\ref{z1}), and we need
to know how to recover it. This problem is addressed in the Lemma below.

\vspace{.3cm}

\noindent
{\bf Lemma 4:} Assume $\Psi(t,x)$ satisfies the wave equation (\ref{z1}) with initial conditions $\Psi(0,x)=: \Psi^0(x)$
 and $\p \Psi / \p t (0,x)=: \dot \Psi^0(x)$.
 Let $\hat \Psi := \I \Psi$,  $\hat \Psi ^0 := \I \Psi ^0$ and $\dot{\hat{\Psi}}^0 := \I  \dot \Psi^0(x)$, then:
 \begin{enumerate}
\item $\hat \Psi$ satisfies the wave equation (\ref{z2}) with
initial conditions $\hat \Psi(0,x)=\hat \Psi ^0$ and $\p \hat \Psi / \p t (0,x)= \dot{\hat{\Psi}}^0$.
\item If $\w_o \neq 0$, $\Psi(t,x)$ can be obtained from $\hat \Psi (t,x)$ by means of
\begin{multline} \label{sol}
 \Psi(t,x) =  \cos (\w_o t) \hat \Psi ^0 + \frac{\sin (\w_o t)}{\w_o} \dot{\hat{\Psi}}^0
+ \frac{1}{w_o} \left(
\sin (\w_o t) \int_0^t \cos (\w_o t') \hat \I \hat \Psi(t',x) dt' -\cos (\w_o t) \int_0^t \sin (\w_o t') \hat \I \hat \Psi(t',x) dt' \right)
\end{multline}
For  $\w_o = 0$ we have
\begin{equation}  \label{sol2}
\Psi(t,x) = \int_0^t \left( \int_0^{t'} \hat \I \hat \Psi(t'',x) dt'' \right) dt' + t \dot{\hat{\Psi}}^0  + \hat \Psi ^0
\end{equation}
\end{enumerate}
\vspace{.3cm}

\noindent
{\bf Proof:} (i) is trivial. To prove (ii) note from Lemma 3, equation (\ref{inverse}),  that
\begin{equation} \label{eqn}
  \hat \I \hat \Psi = \hat \I \I \Psi =  (\w_o^2 - {\cal H}) \Psi = (\w_o^2 + \p^2  / \p t^2) \Psi ,
\end{equation}
where we have used that $\Psi$ satisfies  (\ref{z1}) in the last equality.
 The solution of (\ref{eqn}), regarded as a  differential equation in $t$  on $\Psi$, is
\begin{equation} \label{sola}
\Psi(t,x) =  \cos (\w_o t) F(x) + \sin (\w_o t) K(x) + \frac{1}{w_o} \left(
\sin (\w_o t) \int_0^t \cos (\w_o t')  \hat \I \hat \Psi (t',x) \, dt' -
\cos (\w_o t) \int_0^t \sin (\w_o t')  \hat \I \hat \Psi(t',x) dt' \right)
\end{equation}
if $\w_0^2 \neq 0$, and
\begin{equation}  \label{sol2a}
\Psi(t,x) = \int_0^t \left( \int_0^{t'}  \hat \I \hat \Psi(t'',x) dt'' \right) dt' + t R(x) + Q(x)
\end{equation}
if $\w_o =0$. The unknown functions of $x$, $F$ and $K$ ($Q$ and $R$), are ``integration constants'' of (\ref{eqn}),
 they contain  the information about $\Psi$ that we have lost
when applying $\hat \I \I$. Fortunately, this information  is just the
initial conditions, since it can readily be seen that $F(x) = \Psi(0,x)= \Psi^0(x) $ and $\w_o K(x)=\p \Psi / \p t (0,x) =
\dot \Psi^0 (x)$ ($Q(x) = \Psi(0,x)= \Psi^0(x), \;R(x)=\p \Psi / \p t (0,x) =
\dot \Psi^0 (x)$) .
This gives (\ref{sol}) from (\ref{sola}), and (\ref{sol2}) from (\ref{sol2a}) $\Box$\\

\subsection{Intertwining operator for the negative mass Zerilli equation} \label{apli}

Let ${\cal H}$ be the Zerilli Hamiltonian, and assume an intertwiner is constructed using a solution of
 ${\cal H} \psi_{\w} = \w^2 \psi_{\w}$. Since generic
solutions of this equation behave as (\ref{crc}) (Lemma 2 in \cite{dg}), there is a chance
that the transformed potential (\ref{vh}) be nonsingular at $r_s$, the singularity of $V$ being removed by $-2g'$,
and this may well be a consequence of
\begin{equation}\label{hpsi1}
\hat \Psi = \Psi_z' - \frac{\psi_{\w}{}'}{\psi_{\w}} \Psi_z
\end{equation}
being a smooth function of the perturbed metric. All these expectations
turn out to be right, at least, if we  use the generalization to
arbitrary harmonic number $\ell$ of the solution of
\begin{equation}\label{zm1}
{\cal H} \psi_0 = 0
\end{equation}
found in \cite{ghi} for $\ell=2$. We will first prove the smoothness of $\hat V$, then that of $\hat \Psi$\\

\subsubsection{Smoothness of $\hat V$}

Given $V$ of the form (\ref{poten1}),  $M < 0$, the transformed  potential is
\begin{equation} \label{zm4}
\hat V = V - 2 (\psi_0' / \psi_0 )'.
\end{equation}
 Let us first consider the behaviour of $\hat V$  at the kinematic singularity $r=r_s$.
  Using the fact that $\psi_0$ is a solution of (\ref{zm1}), and turning to $r$ (instead of $x$) derivatives, we find,
\begin{equation}\label{zm5}
\hat{V} = \frac{2(r-2M)^2}{r^2 \psi_0^2}\left(\frac{d \psi_0}{dr}\right)^2 -V(r)
\end{equation}
Now, if $\psi(r)$ is {\em any} solution of (\ref{zm1}),
\begin{equation}\label{zm6}
\psi(r) = a_0 (r-r_s)^{-1}+\frac{a_0(\ell+2)^2(\ell-1)^2}{12 M(\ell^2+\ell+1)}
+a_3(r-r_s)^2+{\cal{O}}\left((r-r_s)^3\right)
\end{equation}
where $a_0$, and $a_3$ are arbitrary constants. Replacing in (\ref{zm5}), assuming $a_0 \neq 0$,
and expanding in powers of $(r-r_s)$, we find,
\begin{equation}\label{zm7}
\hat V = \frac{(\ell^2+\ell+2)(\ell+2)^3(\ell-1)^3}{216 M^2} +\left[\frac{(\ell^2+\ell+1)(\ell+2)^4(\ell-1)^4}{648 M^3}-\frac{4 (\ell^2+\ell+1)^2 a_3}{3 a_0}\right](r-r_s)+{\cal{O}}\left((r-r_s)^2\right)
\end{equation}
which shows that $\hat V$ is smooth for $r=r_s$, provided $a_0 \neq 0$ (if $a_0=0$, $\hat V$ has a second order
pole at $r_s$.) We consider therefore, $\psi_0$ of the form,
\begin{equation}\label{zm2}
\psi_0=\frac{\chi(r)}{6 M+r(\ell+2)(\ell-1)}
\end{equation}
with $\chi$ smooth in $ r \geq 0$. Replacing (\ref{zm2}) in (\ref{zm1}), we find that $\chi$ satisfies,
\begin{equation}\label{zm8}
\frac{d^2 \chi}{dr^2} +\frac{\left[6M^2+2 r \lambda(3M-r)\right]}{r(r-2M)(3M+\lambda r)}\frac{d \chi}{dr}
-\frac{\left[6M^2+2 r \lambda(3M+\lambda r)\right]}{r^2(r-2M)(3M+\lambda r)}\  \chi =0,
\end{equation}
then
\begin{equation}\label{zm5a}
\hat{V} = \frac{2(r-2M)^2}{r^2 \chi^2}\left(\frac{d\chi}{dr}\right)^2
-\frac{4(r-2M)^2 \lambda}{r^2 (3M+\lambda r) \chi}\frac{d\chi}{dr}-
\frac{(r-2M)(6M^2+\lambda r (2\lambda r+4M))} {r^4 (3M+\lambda r)}
\end{equation}
is smooth at $r=r_s$ if $\chi$ is smooth.
The only remaining possible
  singularities for $r >0$ would correspond to the zeros of $\chi$ for $r>0$, since $V$ is smooth except at $r_s$.
   It turns out that (\ref{zm8}) admits, for every $\ell \geq 2$, a polynomial solution of the form,
\begin{equation} \label{zm3}
\chi(r)=\sum_{n=1}^{\ell+2}\frac{(n-2)\left[(n-4)\ell(\ell+1)+n-1\right]\Gamma(\ell+n-1)
}{ 2^{n}\Gamma(n)^2\Gamma(\ell-n+3) (-M)^{n}}r^n ,
\end{equation}
which, for $\ell=2$ reduces to the solution (\ref{l2zm}) found in \cite{ghi},
\begin{equation} \label{zm3a}
\chi(r)=-\frac{3r}{2M}+\frac{3r^3}{4M^3}+\frac{r^4}{4M^4}
\end{equation}
which is positive for $M<0$ and $r>0$. Similarly, for $M<0$, and $\ell \geq 3$ we have,
\begin{eqnarray} \label{zm3b}
\chi(r) & = & \frac{r}{|M|} \left[\frac{3}{2}-\frac{\ell(\ell+2)(\ell^2-1)}{32 } \frac{r^2}{|M|^2}+\frac{\ell(\ell+2)(\ell^2-1)}{96}\frac{ r^3}{|M|^3} \right.
\nonumber \\ & &\left. +
\frac{\ell(\ell^2+\ell+4)(\ell+3)(\ell^2-4)(\ell^2-1)}{6144}\frac{ r^4}{|M|^4}+\dots\right]
\end{eqnarray}
where all the remaining terms, indicated by dots, are non negative for $r>0$. The fourth degree polynomial given explicitly between the brackets in (\ref{zm3b}) is positive for $r=0$ and  for sufficiently large $r$. Therefore, it can only have a zero if its derivative vanishes at least at one point for $r>0$. One can check that for $r > 0$ there is only one root given by,
\begin{equation}\label{zm9}
r_0 = \frac{4|M|\left(\sqrt{6\ell(\ell^2-1)(\ell+2)-108}-6\right)}{(\ell-2)(\ell+3)(\ell^2+\ell+4)}.
\end{equation}
This must correspond to a minimum of the polynomial in $r > 0$.
Replacing $r=r_0$ in (\ref{zm3b}) we find,
\begin{equation}\label{zm10}
\chi(r_0) \geq \frac{16 (\rho-6)\left((\ell^3-\ell)(\ell+2)-18\right)\left((\ell^3-\ell)(\ell+2)(\rho-18)+288\right)}
{(\ell^2+\ell+4)^4(\ell-2)^4(\ell+3)^4}
\end{equation}
where $\rho=\sqrt{6\ell(\ell^2-1)(\ell+2)-108}$. The right hand side of (\ref{zm10}) is positive for $\ell \geq 3$. We conclude
 that $\chi(r) > 0$  for $r >0$. This completes  the proof of the smoothness of $\hat V$. \\
The explicit form of $\hat V$ as a function of $r$ for generic $\ell>2$
is very complicated but, fortunately, it is not required for the rest of our analysis.
In any case, it is possible to obtain several features of $\hat V$ directly from  (\ref{zm5a}). First, since $\chi$ is a polynomial of degree $\ell+2$,  we find that for large $r$,
\begin{equation}\label{zm11a}
\hat V = \frac{(\ell+2)(\ell+1)}{r^2} + {\cal{O}}(r^{-3}) > 0.
\end{equation}
 Also,
from (\ref{zm3}), for $r \to 0$ we have, in general
\begin{equation}\label{zm11}
\hat V = 12 M^2 r^{-4} -2 M (\ell^2+\ell+3) r^{-3} +{\cal{O}}(r^{-2})
= \frac{3}{4 x^2} +\frac{\ell^2+\ell-1}{4|M|^{1/2}x^{3/2}} -\frac{\ell(\ell+1)}{4 |M|x} +{\cal{O}}(x^{-1/2})
\end{equation}
Thus the general local solution of the differential equation $\hat {\cal H} \hat \psi_E = E \hat \psi_E$, for $x \to 0^+$ is of the form,
\begin{equation}
\hat \psi_E = a_0 \left( x^{3/2} + ... \right)  + b_0 \left(  x^{-1/2} + ... \right)
\end{equation}
which is not square integrable near $x=0$ unless $b_0=0$. This last condition
can easily be checked to correspond
 precisely to the $\theta=0$ boundary condition for the local solutions of ${\cal H} \Psi \propto \Psi$ in (\ref{sae}).

\subsubsection{Smoothness of $\hat \psi$}

We need only check smoothness at $r_s$, and this follows from  equations (\ref{zm5}) and (\ref{crc}),
 which imply that $\hat \psi = \I \psi_z$ admits a Taylor
expansion around $r=r_s$.
Of particular relevance is the transformed of $\I \psi^{unst}$ which, from the above results, belongs to ${\cal D}$ and thus is a
negative energy eigenfunction of $\hat {\cal H}$, which, therefore, has, at least, one bound state.
This, by the way, implies that $\hat V$ must have a region where it takes negative values, as can be explicitly checked for particular
 values of $\ell$, and is illustrated in Figure 2 (right panel).\\

\subsection{Intertwining operator for the positive mass Zerilli equation}

\subsubsection{ Smoothness of $\hat V$}

The intertwining transformation is equally applicable when $M>0$. One can check that equations
(\ref{zm5}) and (\ref{zm2})-(\ref{zm3}) are still valid if $M>0$. From (\ref{zm5a}) we see that for $r \geq 2M$, the only possible singularities of $\hat{V}$ correspond to zeros of $\chi(r)$, then we need to prove that $\chi(r)$ has no zeros in $r \geq 2M$.
This can be seen as follows: first we notice that near $r=2M$, (\ref{zm8}) has only one regular solution, and this must correspond to the
polynomial solution (\ref{zm3}). Expanding this solution in powers of $(r-2M)$ we find,
\begin{equation}
\label{posM01}
\chi \left( r \right) ={  a_0}
+{\frac {  \left( 4{
\lambda}^{2}+6\lambda+3 \right) {a_0}   }{ 2\left( 2
\lambda+3 \right) M}}\left( r-2M \right)
+{\frac {\lambda  \left( \lambda+
1 \right) {a_0} }{4 {M}^{2}}} \left( r-2M \right) ^{2}
+{\cal{O}} \left( (r-2M ) ^{3}\right)
\end{equation}
where $a_0$ is a constant. This implies that $\chi$ and $d\chi/dr$ are both non vanishing and have the same sign and, therefore, $\chi$ is increasing if $\chi(r=2M)>0$,  decreasing if $\chi(r=2M)<0$. Then, in order to have a zero for $r>2M$, there must be a point where $d\chi/dr=0$. But from \ref{zm5a} we notice that for $r>2M$, at any point where $d\chi/dr=0$ we must have $\chi$, and $d^2\chi/dr^2$ with the
 same sign, namely, this corresponds to a minimum for positive $\chi$ and a maximum for negative $\chi$. But since, e.g., for $\chi(r=2M) >0$, the function is already increasing, and the condition $d\chi/dr=0$ cannot be satisfied for $r >2M$,
  implying that $\chi(r)$ has no zeros for $r \geq 2M$, and similarly for $\chi(r=2M)<0$. Thus $\hat{V}(r)$ is regular for $r\geq 2M$,
   it vanishes as $(r-2M)$ for $r \to 2M$ (see  (\ref{zm5a})), and as $1/r^{(\ell+2)(\ell+1)}$ for large $r$. In this respect, it is similar to the Zerilli potential $V(r)$. One can see, however, that $\hat{V}(r)$ is not positive definite, (see Figure 2, left panel for an example), making the proof of the stability of the exterior region of a Schwarzschild black hole more complicated in the context of the $\hat \Psi$ formulation.

\subsubsection{ Smoothness of $\hat \Psi$}

The proof for $M<0$ holds also for positive mass.

\subsection{Proof of the Theorem}

Parts (i), (ii) and (iii) of the Theorem were proved in the two previous subsections. Part (iv) follows from Lemma 4 and a
uniqueness argument: since there is a unique solution in ${\cal D}$ of the equation $\hat {\cal H} \hat \Psi =0$
with initial condition $(\hat \Psi ^0, \dot{\hat{\Psi}}^0) \in {\cal D}  \times {\cal D}$, and $\I \Psi_z$ is such a solution if $\Psi_z$
solves Zerilli's equation with initial data $(\Psi^0, \dot \Psi^0)$ and boundary condition $\Psi_z \simeq x^{1/2}$
for $x \simeq 0$, it must be $\hat \Psi =\I \Psi_z$, then part (iv) follows from Lemma 4 and the fact that $\w_o=0$
for the intertwiner
$\I$ that we use.

\section{Summary} \label{sum}

The propagation of gravitational perturbations on a negative mass Schwarzschild background is a subtle
problem for two reasons. First, this space is not globally hyperbolic. As a consequence,
the perturbation equations can be reduced to a single $1+1$ wave equation with a space dependent potential
for the so called Zerilli function, restricted to a
semi-infinite domain $x>0$, $(t,x)$ being standard inertial coordinates on two dimensional Minkowski space,
$x=0$ the position of the singularity.
This implies that a physically motivated choice of boundary conditions
at $x=0$ is required. There is a unique choice dictated simultaneously by two conditions \cite{ghi,dg}: (i) that
the linearized regime be valid in the whole domain, and, in particular, that the invariants made out of the Riemman tensor
behave such that their first order piece does not diverge faster than their zeroth order piece as the singularity is approached;
 (ii) that the energy of
the perturbation, as measured using the second order correction to the Einstein tensor \cite{ghi} be finite.\\
The second problematic issue with the standard approach is not essential, but related to a choice of variables:
  the Zerilli function $\Psi_z$ is a singular function of the first order metric coefficients. As a consequence, the wave
equation it obeys has  a potential  with a ``kinematic'' singularity, and it is not clear how to evolve
initial data, since the usual approach of separation of variables leading to a well behaved quantum Hamiltonian operator
for the $x$ coordinate breaks down.\\
We have introduced an alternative diagonalization of the linearized even mode Einstein's equation around a Schwarzschild spacetime,
 using a field $\hat \Psi$
 which is smooth for regular metric perturbations, regardless the sign of the mass $M$. This field obeys a wave equation
with a smooth potential, that can be solved by separation of variables.  Moreover,
the spatial piece of the modified wave equation has a unique self adjoint extension, that naturally selects the boundary
condition that is physically relevant.\\
The connection between the two fields is provided by an intertwining operator,
 $\hat \Psi = \Psi_z - \psi_0'/\psi_0 \Psi_z =: {\cal I}
\Psi_z$, similar to the operators linking supersymmetric pairs of quantum hamiltonians.
We have also shown that, in spite of the fact that ${\cal I}$ has a non trivial kernel, it is possible to
evolve the perturbation equations using, at two different steps, the initial condition for the Zerilli function.
A straightforward application of this formalism allows us to show that the unstable mode
found in \cite{dg} can actually be excited by initial data compactly supported away from the singularity. This
closes a gap in our proof in \cite{dg} of the linear instability of the negative mass Schwarzschild spacetime.

\section*{Acknowledgments}

We thank Gast\'on \'Avila and Sergio Dain for useful comments on the manuscript.
This work was supported in part by grants from CONICET (Argentina)
and Universidad Nacional de C\'ordoba.  RJG and GD are supported by
CONICET.

\end{document}